\documentclass{JHEP3}

\usepackage{epsfig}
\usepackage{amsfonts,amsmath,amssymb,amsthm,amstext,eucal,array}
\usepackage{graphicx,lscape}

\def\CC{{\cal C}}

\def\CK{{\cal K}}
\def\CL{{\cal L}}

\def\ta{\tilde{a}}
\def\tc{\tilde{c}}
\def\tb{\tilde{b}}

\newcommand{\bR}[0]{{\bf R}}

\newcommand{\ads}[1]{{\rm AdS}_{#1}}

\newcommand{\sph}[1]{{\rm S}^{#1}}

\newcommand{\be}{\begin{equation}}
\newcommand{\ee}{\end{equation}}
\def\bea{\begin{eqnarray}}
\def\eea{\end{eqnarray}}

\def\tr{\hbox{tr}}

\newcommand{\dvol}{\text{dvol}}
\newcommand{\fso}{\mathfrak{so}}

\newcommand{\Og}{\mathrm{O}}
\newcommand{\SO}{\mathrm{SO}}

\newcommand{\gU}{\mathrm{U}}
\newcommand{\SL}{\mathrm{SL}}
\newcommand{\fsl}{\mathfrak{sl}}
\newcommand{\U}{\mathrm{U}}

\font\mybb=msbm10 at 10pt
\def\bb#1{\hbox{\mybb#1}}
\def\bR {\bb{R}}
\def\bZ {\bb{Z}}

\def\bI {\bb{I}}

\def\bT {\bb{T}}

\preprint{hep-th/0311237 \\
ITFA-2003-52 \\ UPR-1055-T\\  WIS/30/03-NOV-DPP}
\title{A Multi-boundary AdS Orbifold and DLCQ Holography: \\ {\Large A 
Universal Holographic Description of Extremal Black Hole Horizons} }
\author{
Vijay Balasubramanian$^1$, Asad Naqvi$^{1,2}$ and Joan Sim\'{o}n$^{1,3}$\\
~\\
1. David Rittenhouse Laboratories,
The University of Pennsylvania,
Philadelphia,  \\ 
~~~~PA 19104-6396, U.S.A.\\
2. Instituut voor Theoretische Fysica, 
Valckenierstraat 65, 1018XE Amsterdam, The Netherlands \\
3. The Weizmann Institute of Science, 
Herzl Street 2, 76100 Rehovot, Israel\\
~\\
\email{vijay@endive.hep.upenn.edu, anaqvi@science.uva.nl, 
jsimon@bokchoy.hep.upenn.edu}
}

\abstract{We examine a stationary but non-static asymptotically $\ads{3}$ spacetime 
with two causally connected conformal boundaries, each of which is a ``null cylinder'', namely a cylinder with a null direction identified.    
This spacetime arises from three different perspectives: 
(i) as a non-singular, causally regular orbifold of global $\ads{3}$ by boosts, (ii) as a Penrose-like limit focusing on the horizon of extremal BTZ black holes, and (iii) as an $S^1$ fibration over $\ads{2}$.    Each of these perspectives sheds an interesting light on holography. 
Examination of the conformal boundary of the spacetime shows that the dual to the space should involve DLCQ limits of the D1-D5 conformal field theory.    The Penrose-like limit approach leads to a similar conclusion, by isolating a sector of the complete D1-D5 CFT that describes the physics in the vicinity of the horizon of an extremal black hole.  As such this is a holographic description of  the universal horizon dynamics of the extremal black holes in $\ads{3}$ and also of the four and five dimensional stringy black holes whose states were counted in string theory.  The $\ads{2}$ perspective draws a connection to a $0+1$d quantum mechanical theory.  Various dualities lead to a Matrix model description of the spacetime.  Many interesting issues that are related to both de Sitter physics and attempts to ``see behind a horizon'' using AdS/CFT 
arise from (a) the presence of two disconnected components to the boundary, and (b) the analytic structure of bulk physics in the complex coordinate plane.
}

\begin{document}

\section{Introduction}

There are several issues concerning string theory in time dependent spacetimes that we would
like to study using holography.  First, we would like to gain insight into the resolution of 
singularities localized in time such as those that occur inside black
holes and at the big bang.  Second, we would like to understand how
and whether physics behind a horizon is represented and whether the
information loss paradox for black holes is avoided.  Third, we would
like to understand how holography and quantum gravity work in the
presence of a positive cosmological constant.  Finally, we would like
to have simple solvable examples of time-dependent universes in string
theory, that avoid  pitfalls such as closed causal curves and large
back reaction that make typical constructions such as boost orbifolds
difficult to define consistently and to understand.\footnote{Recent
relevant work includes \cite{lororb,lororb2}.}   In many of the cases
that have been studied, such as the eternal black hole
\cite{eternalBH, BKLT, eskoper,BTZholog} and de Sitter space
\cite{dsdual}, one of the challenges that arises is that the spacetime
has multiple disconnected components to its  conformal
boundary\footnote{Interesting prior work exploring the entropy of dS
black holes from the perspective of the Chern-Simmons approach to 3d
gravity appears in \cite{muinpark}.}.    
If holographic duals are to be associated with the conformal boundary of a space it is necessary to understand how the theories associated each disconnected component are related.

In this paper we study an orbifold of $\ads{3}$ by two boosts which does not have closed causal curves. In terms of $\ads{3}$ as an $SL(2,R)$ group manifold, the orbifold is an identification by a hyperbolic element of the left $SL(2,R)$ action.   The orbifold is closely related to the extremal BTZ \cite{BHTZ} spacetimes, but is nevertheless not a black hole -- there is neither a horizon nor a singularity.   However, like an eternal black hole and like de Sitter space, the spacetime has two disconnected components to its conformal boundary.  Unlike the former cases the two boundary components are causally connected. At first sight this appears to violate a no go theorem due to Galloway et al.~\cite{galloway} which says that in three dimensions or higher, any Lorentzian AdS spacetimes with multiple boundaries also have horizons.   However, as we will see, our 3d orbifold contains a circle of fixed size.  Compactifying on this circle gives rise to a 2d AdS geometry with flux which is outside the conditions required for the theorem of~\cite{galloway}.

Each of the boundary components is a ``null cylinder", namely a cylinder in which a lightlike direction has been identified, or equivalently the infinite momentum boost of the usual cylinder with a spacelike identification.  Indeed, the timelike surfaces at fixed radial coordinate are literally conformal to boosted cylinders with the boost approaching infinity as the spacetime boundary is reached.   The space, which has appeared before as the ``self-dual'' $\ads{3}$ geometry of Coussaert and Henneaux \cite{CH94},  can be understood in many ways.   It also 
arises from a Penrose-like limit focusing on the horizon of extremal BTZ black holes \cite{lowe, S98,ads2},  as discussed by Strominger and collaborators.   Thus it is the universal description of the vicinity of the horizon in all of the extremal finite-area black holes whose entropy has been explained in string theory.  Finally, the orbifold can  be seen as a circle fibration of $\ads{2}$ or, equivalently, $\ads{2}$ with a constant flux turned on.

To study holography in the orbifold the first step is to extract the normalizable and non-normalizable fluctuations of bulk field theory to assemble a holographic dictionary and prescription for computing correlations.   By exploring unitary representations of the $SL(2,R) \times U(1)$ isometry group that survives the orbifold projection we find a quantized spectrum of normalizable modes corresponding to states in the dual field theory \cite{BKL}.    Modes with positive (negative) angular momentum are localized closer to the right (left) boundary of the spacetime.    From the perspective of directly solving the wave equation on the orbifold the same spectrum is obtained by requiring single valuedness of wavefunctions in the complex coordinate plane.  This requirement of an absence of cuts in the complex plane is reminiscent of other recent work in which the structure of amplitudes and wavefunctions in the complex plane was important for the structure of holography in the BTZ black hole background \cite{eskoper,BTZholog}. 

We also obtain the non-normalizable modes that correspond to sources in the dual field theory \cite{BKL}.    Again requiring the absence of cuts in the complex coordinate plane
requires a particular addition of normalizable modes to the non-normalizable basis, thus choosing a distinguished  vacuum.
The non-normalizable  modes diverge on both boundaries simultaneously suggesting that it is not possible to turn on  independent sources for components of the dual that are associated with each boundary.  However, on closer examination it turns out that in fact the correlation is between sources at real positions on one boundary and shifted in the complex coordinate plane in the other.   Boundary data, and thus dual sources, at real positions are in fact independent.   Indeed, it turns out that the non-normalizable modes in the BTZ black hole also diverge at both boundaries in a similar way.  Since there are two disconnected, but entangled, components to the dual in that case, a similar picture is suggested in our case also.   The standard calculation of CFT 2-point correlators from the bulk on-shell action \cite{GKPW} takes contributions from both boundaries.  However, by expressing the calculation in terms of data at one or both boundaries we can compute correlation function on each boundary component or between them.    We also compute a bulk-boundary propagator on the orbifold by calculating the bulk Feynman propagator from a sum on images and then taking a bulk point to the boundary.  From this perspective, the dual two-point function arises from taking both bulk points to a boundary; so it is possible to compute separate correlators in each dual component, as well as between them.   The results match the on-shell calculation, and we find that the correlation function between boundaries is obtained from the one within a boundary by certain shifts of the arguments in the complex coordinate plane.  This situation, involving analytic relations between different correlators via excursions in the complex coordinate plane is strongly reminiscent of the holography in the BTZ black hole \cite{eskoper,BTZholog}.

The various perspectives from which our spacetime arises (as an orbifold of $\ads{3}$, as a space with two null cylinder boundaries, as an $\ads{2}$ fibration, and as a Penrose-like limit of the D1-D5 system) give us a number of tools for studying the dual field theory.   Recall first that the dual to global $\ads{3}$ spacetime is the D1-D5 CFT, namely a
sigma model on the target space $K3^N/S_N$ (or $T4^N/S_N$ depending on how $\ads{3}$ is embedded in string theory).  Then, as an orbifold of $\ads{3}$, the dual theory should be an orbifold of the 
D1-D5 CFT by a certain left-moving conformal transformation.  States that survive the orbifold are characterized by their right-moving $SL(2,R)$ representation and an integer $U(1)$ charge from the left-moving side.  Of course these surviving symmetries match the isometries of the bulk orbifold.     As a space with two null cylinder boundaries in causal contact, our orbifold should be dual to two  CFTs each of which is Discrete Light Cone Quantized (DLCQ).   Finite energy states in a DLCQ CFT  carry momenta in only the left-moving (or right-moving)  direction.   The two boundaries in our orbifold each produce a tower of positive (negative) integer momenta which together match the complete tower of states in the conformal-orbifold picture.  
There is also a nice correspondence with the fact that normalizable modes in the bulk are localized closer to one boundary or the other depending on the direction in which they rotate, or equivalently the sign of their $U(1)$ charge.    The origin of our spacetime in a Penrose-like limit focusing on the horizons of extremal BTZ black holes allows us to isolate a sector of the D1-D5 CFT that describes this region --  an entangled state 
of the DLCQ of the D1-D5 string makes an appearance.   The perspective that the spacetime is an $\ads{2}$ fibration suggests that our orbifold should be related to a $0+1$d quantum mechanics.  Finally, various S and T dualities, combined with lifts to M-theory lead to  Matrix model descriptions of our spacetime.

\section{Classical geometry}
\label{sec:setup}

Three dimensional anti-de Sitter space $(AdS_3)$ is a maximally symmetric
space of constant negative curvature. It is the hyperboloid
\begin{equation}
  \begin{aligned}
    \text{AdS}_3 & \hookrightarrow \bR^{2,2} \\
    -u^2 - v^2  & + x^2 + y^2 = -l^2 \quad ,
  \end{aligned}
 \label{embedding}
\end{equation}
in flat $\bR^{2,2}$. By construction, the isometry group is $\SO(2,2)$ \footnote{The isometry group of the geometry is really $\Og(2,2)$ but when embedded in string theory, 
the presence of fluxes restricts this to $\SO(2,2)$. }.

A global parametrization of $\ads{3}$ is obtained by solving \eqref{embedding} in terms of
\begin{equation}
  \begin{aligned}
    u&=l\,\cosh\rho\,\cos\tau \quad , \quad
    v=l\,\cosh\rho\,\sin\tau \\
    x &=l\,\sinh\rho\,\cos\theta \quad , \quad
    y = l\,\sinh\rho\,\sin\theta~.
  \end{aligned}
 \label{eq:global}
\end{equation}
The induced metric is
\begin{equation}
  g_{AdS_{3}} = l^2\left[-(\cosh\rho)^2 (d\tau)^2 + (d\rho)^2 +
  (\sinh\rho)^2 (d\theta)^2\right].
 \label{eq:gmetric}
\end{equation}
As usual, we shall refer to $\text{AdS}_3$ as the
universal covering space of the above hyperboloid \eqref{embedding} in which
the global timelike coordinate $\tau$ has been unwrapped (i.e. take
$-\infty < \tau < \infty$).
The Killing vectors of the metric generate the Lie algebra  $\fso(2,2)$ of the isometry group, and 
are described in terms of the embedding space $\bR^{2,2}$ as
\begin{equation}
  J_{ab}=x_b \partial_a -x_a \partial_b ~,
\label{bulksl2gens1}
\end{equation}
with $x^a \equiv (u,v,x,y)$ and $x_a=\eta_{ab}\,x^b$, with 
$\eta_{ab}=(-\,,-\,,+\,,+)$. We can decompose $\fso(2,2)=\fsl(2,\bR)\oplus\fsl(2,\bR)$ via the  linear combinations
\begin{equation}
  \xi_{1}^{\pm} = \frac{1}{2}(J_{01} \pm J_{23}) ~~;~~
  \xi_{2}^{\pm} = \frac{1}{2}(J_{02} \pm J_{13}) ~~;~~
  \xi_{3}^{\pm} = \frac{1}{2}(J_{03} \mp J_{12}) \, ,
\end{equation}
where $\xi_{i}^\pm$ satisfy the non-vanishing commutation relations
\begin{equation}
  [\xi^{\pm}_{i}, \xi^{\pm}_j] = \epsilon_{ijk} \, \xi^{\pm}_k \, .
\label{bulksl2gens2}
\end{equation}

Discrete quotients of $\ads{3}$ involving a single generator are classified by the set of inequivalent uniparametric subgroups of $\SO(2,2)$ \cite{BHTZ,joanjose,ross}.  We will consider quotients by the action of a subgroup of $\SO(2,2)$ isomorphic to $\bZ$.
\begin{equation*}
  \text{P} \rightarrow e^{t\xi} \text{P}~~~,~~~t=0,
  \pm 2\pi, \pm 4 \pi, \cdots \quad \forall\,\text{P}\in\text{AdS}_3, 
 \label{orbact}
\end{equation*}
where
\begin{equation}
  \xi = {1\over 2}\left(J_{02} + J_{13}\right)~.
 \label{eq:killvect}
\end{equation}
This generator is a linear combination of a boost in the ux-plane and
vy-plane in the embedding space $\bR^{2,2}$.

It is instructive to compare the space obtained after this identification with
the BTZ black hole. The latter is obtained by identifying $\ads{3}$ by 
the discrete action generated by the Killing vector \cite{BHTZ}
\begin{equation}
  \xi_{\rm BTZ}={r_+ \over l}J_{12}-{r_- \over l}J_{03}-J_{13}+J_{23}.
\label{nonextident}
\end{equation}
In the non-extremal case, $r_+^2 -r_-^2 >0$ and by a $\SO(2,2)$ 
transformation, $\xi_{\rm BTZ}$ can be brought into the form:
\begin{equation}
  \xi_{\rm BTZ}^\prime={r_+ \over l}J_{12}-{r_- \over l}J_{03}.
\label{xiprime}
\end{equation}
The mass and angular momentum of the black hole are given by
\begin{equation}
  M={1 \over l^2}(r_+^2+r_-^2)\quad , \quad J={2 \over l} r_+r_-.
\end{equation}
The extremal black hole is obtained by taking the limit $r_+ \to r_-$ in (\ref{nonextident}), so that  the generator becomes
\begin{equation}
 \xi_{BTZ} \to \frac{r_+}{l}\left(J_{12} - J_{03}\right) - J_{13} + J_{23}~.
\end{equation}
After a rotation in the $\{x^2,\,x^3\}$ plane by $\pi/2$ this vector field becomes 
\begin{equation}
  \xi_{BTZ} \to \frac{r_+}{l}2\xi + J_{12} + J_{23}~.
 \label{eq:limit}
\end{equation}
where $\xi$ is the generator (\ref{eq:killvect}) of our orbifold spacetime.  
Thus the extremal BTZ identification differs from the one generating our spacetime by the
extra action generated by $J_{12} + J_{23}$. 
The alert reader might worry
that the generator $\xi_{{\rm BTZ}}^\prime$ which is a $SO(2,2)$ rotation 
of $\xi_{{\rm BTZ}}$ does appear to approach
our orbifold generator after a further rotation in the $\{x^2,\,x^3\}$ plane when we take the extremal limit 
$r_- \to r_+$.  However, this is misleading -- the $SO(2,2)$ transformation relating  (\ref{nonextident}) to
(\ref{xiprime}) does not exist in the $r_- \to r_+$ limit.
Indeed, the generator
of the extremal BTZ black hole lies in a different orbit of $SO(2,2)$ than the generator of the self-dual
orbifold.

\subsection{Bulk geometry}
\label{sec:bulk}

By construction, any discrete quotient of $\ads{3}$ is locally isometric
to $\ads{3}$, but differs from it globally. A unique feature
 of \eqref{eq:killvect} is that it is the
only generator giving rise to a smooth discrete quotient of $\ads{3}$
preserving one half of the supersymmetries. This feature is inherited by
any higher dimensional AdS spaces.
Here, by smoothness, we mean absence of fixed points and closed causal curves.
That there are no fixed points can be seen from  the
point of view of the embedding space $\bR^{2,2}$, where the only fixed point is
the origin  $(u=v=x=y=0)$, since Lorentz transformations act linearly
in $\bR^{2,2}$. However,  the origin does not belong to $\ads{3}$; hence the absence 
of fixed points.

To show the absence of closed causal curves, and also to gain some insight
concerning the geometry of the identifications, it is convenient to parametrize
$\ads{3}$ in the  adapted coordinates \cite{CH94}:
\begin{equation}
  \begin{aligned}
    u & = l\left(\cosh z \cosh \phi \cos t + \sinh z \sinh \phi \sin t
    \right),  \\
    v & = l\left(\cosh z \cosh \phi \sin t - \sinh z \sinh \phi \cos t
    \right), \\
    x & = l\left(\cosh z \sinh \phi \cos t + \sinh z \cosh \phi \sin t
    \right), \\
    y & = l\left(\cosh z \sinh \phi \sin t - \sinh z \cosh \phi \cos t
    \right)  \,.
  \end{aligned}
 \label{eq:adapted}
\end{equation}
In these coordinates, the $\ads{3}$ metric is
\begin{equation}
  g=l^2\left(-(dt)^2+(d\phi)^2+2 \sinh(2z) dt d\phi + (dz)^2\right) \, ,
 \label{eq:fmetric}
\end{equation}
and, before making any identification, all coordinates are taken to
be non-compact, $-\infty < t\,,\phi\,,z < \infty$.

Since the action of the generator \eqref{eq:killvect} in the adapted coordinate
system \eqref{eq:adapted} is given by a simple shift along the $\phi$
direction,
\[
  2\xi = \frac{\partial}{\partial\,\phi}\,,
\]
the description of the discrete quotient is given by making $\phi$ an angular coordinate taking values in  $[0\,,2\pi)$.   Notice that the effective radius of the circle is constant and equal to the radius of $\ads{3}$,
$l$.

It is now easy to prove the non-existence of closed causal curves \cite{CH94}.
We begin by assuming that such a closed causal curve exists. Then, if
 $x^\mu (\lambda)$ is the embedding of this curve with $\lambda$ being its affine
parameter, the norm of its tangent vector satisfies the condition
\begin{equation*}
  -\left(\frac{dt}{d\lambda}\right)^2 + \left(\frac{dz}{d\lambda}\right)^2
  + \left(\frac{d\phi}{d\lambda}\right)^2 + 2\sinh 2z \frac{dt}{d\lambda}
  \frac{d\phi}{d\lambda} \leq 0 \quad \forall \lambda .
\end{equation*}
The only way in which the causal curve can be closed is by connecting
any initial point $(t_0\,,z_0\,,\phi_0)$ with $(t_0\,,z_0\,,\phi_0+\Delta)$.
This automatically requires the existence of, at least one value of
the affine parameter $\lambda$, say $\lambda=\lambda_\star$, where
the timelike component of the tangent vector to the causal curve
vanishes
\begin{equation*}
  \exists \lambda=\lambda_\star \quad \text{s.t.} \quad
  \left.\frac{dt}{d\lambda}\right|_{\lambda_\star} = 0 ~.
\end{equation*}
It is clear that the only way to satisfy the causal character of the curve
at $\lambda_\star$ is by having a vanishing tangent vector at that point,
which contradicts the assumption of $\lambda$ being an affine parameter.
Thus we conclude that our discrete quotient is free of  closed causal curves.

In order to identify the isometries of the quotient manifold, it is 
useful to describe $\ads{3}$ as the $\SL(2,\bR)$ group manifold in terms
of $2\times 2$ matrices. An explicit global parametrization for this
group manifold adapted to the action of the discrete quotient is given by
\begin{equation}
  \hat{g}=e^{\phi\sigma_1}~ e^{z\sigma_3}~ e^{it \sigma_2}\,,
\end{equation}
where $\sigma_i$ are the standard Pauli matrices. Then,  the metric
in adapted coordinates \eqref{eq:fmetric} can be written as
\begin{equation*}
  g =\frac{l^2}{2} \tr (\hat{g}^{-1}d\hat{g})^2.
\end{equation*}
The isometry group is  $\SL(2,\bR) \times \SL(2,\bR)$, and its action
is given by left and right multiplication:
\begin{equation}
  (h_L,h_R)\in \SL(2,\bR) \times \SL(2, \bR): ~ \hat{g} \rightarrow h_L ~
  \hat{g} ~h_R ~.
\end{equation}
In this formulation,  the discrete quotient under discussion
is  the identification
\begin{equation}
  \hat{g} \sim e^{2 \pi \sigma_1} \hat{g} \,,
 \label{eq:ident}
\end{equation}
which implies that $\phi \sim \phi+2\pi$. Notice that the identification is by an action of a hyperbolic element in the left $SL(2,R)$ and the trivial element of the right $SL(2,R)$.

The isometries of the quotient manifold are given in terms
of the  generators that commute with the action of the discrete quotient.
Since the latter does not act on the right $\SL(2,\bR)$ factor, the
isometry group will contain $\SL(2,\bR)$ besides the action along the quotient
itself.  Thus  the isometries of the background \eqref{eq:fmetric}
generate a $\gU(1)\times\SL(2,\bR)$ group. Its generators, in the adapted
coordinate system \eqref{eq:adapted}, are given in terms of the Killing
vectors
\begin{equation}
  \begin{aligned}
    \xi &= \frac{1}{2} \frac{\partial}{\partial \phi}\,, \\
    \chi_1 &= \frac{1}{2} \frac{\partial}{\partial t} \,, \\
    \chi_2 & = \frac{1}{2} \tanh(2z)\cos(2t) \frac{\partial}{\partial t}
    + \frac{\cos{2t}}{2\cosh(2z)}\frac{\partial}{\partial \phi} +
    \frac{1}{2} \sin(2t) \frac{\partial}{\partial z}\,, \\
    \chi_3 & =  -\frac{1}{2}\tanh(2z)\sin(2t) \frac{\partial}{\partial t}-
    \frac{\sin{2t}}{2\cosh(2z)}\frac{\partial}{\partial \phi} +
    \frac{1}{2} \cos(2t) \frac{\partial}{\partial z} \,,
  \end{aligned}
\label{isometries}
\end{equation}
where $\{\chi_i=\xi_i^{-}\}$ satisfy the $\fsl(2,\bR)$ commutation relations
\[
  [\chi_1\,,\chi_2] = \chi_3 \quad, \quad [\chi_3\,,\chi_1] = \chi_2
  \quad ,\quad [\chi_2\,,\chi_3] = -\chi_1~.
\]

\paragraph{Compactification to $\ads{2}$: } The existence of an $\SL(2,\,\bR)$ isometry group suggests a close relationship
between our orbifold \eqref{eq:fmetric} and $\ads2$. This relation can be made
precise by realizing that the metric \eqref{eq:fmetric} is an $S^1$ fibration
over $\ads2$. Indeed, we can rewrite \eqref{eq:fmetric} as
\begin{equation}
  g = l^2\left(-\cosh^2(2z)\,dt^2 + dz^2 + \left(d\phi + \sinh (2z)\,dt\right)^2\right)~.
 \label{eq:fibration}
\end{equation}
Compactifying on $\phi$ now gives the metric 
\begin{equation}
  \begin{aligned}[m]
    g_2 &= -\cosh^2 2z \, dt^2 + dz^2~, \\
    A_1 &=  \sinh 2z \,dt .
\end{aligned}
\end{equation}
The metric is precisely that of $\ads{2}$, but there is also a constant electric field.\footnote{The field is constant in the sense that the field strength of the $\U(1)$ connection is proportional to the $\ads{2}$ volume form.}  As we will see in 
Sec.~\ref{sec:confbound}, the two 
disconnected conformal boundaries of $\ads{2}$ are also reflected in the boundary of the 3d space.

\paragraph{Supersymmetry: } When considering the above construction in a supergravity context, it is 
important to analyze the supersymmetry preserved by the discrete identification.
This was already discussed in \cite{CH94} by explicit computation of the Killing 
spinors in the adapted coordinates \eqref{eq:adapted}.   The conclusion of that paper was that the configuration
\eqref{eq:fmetric} preserves one half of the spacetime supersymmetry.
A much simpler way to get to the same conclusion is, once more, to think of
$\ads{3}$ as embedded in $\bR^{2,2}$. The following discussion is based on
the general discussion that will be presented in \cite{joanjose}.

Killing spinors in this embedding space are just constant spinors $\varepsilon_0$ and they have two
different chiralities. When decomposing the type IIB chiral spinors into tensor products of spinors in $\ads{3}$,
$S^3$ and a 4-torus, the Majorana spinors in $\ads{3}$ transform in two different representations.
Each of these is mapped to a different chiral sector in $\bR^{2,2}$. 
The amount of supersymmetry preserved by the discrete identification is obtained by analyzing how many of the constant
parallel spinors $\varepsilon_0$ remain invariant under the action of the generator of the discrete group
\eqref{eq:killvect}. Infinitesimally, the latter condition is equivalent to \cite{Kosmann,JMFKilling}
\begin{equation}
  L_\xi\varepsilon_0 \equiv \nabla_\xi\varepsilon_0 + \frac{1}{4}\partial_{[m}
  \xi_{n]}\Gamma^{mn}\,\varepsilon_0 = 0~,
\end{equation}
which gives rise to the algebraic constraint
\begin{equation}
  \Gamma_{uvxy}\,\varepsilon_0 = \varepsilon_0 ~,
\end{equation}
which is only satisfied by half of the components of the Killing spinor,
since $\left(\Gamma_{uvxy}\right)^2=\bI$ and $\text{Tr}\, \Gamma_{uvxy}=0$.

The above condition tells us that only the subset of parallel spinors in $\bR^{2,2}$ with a positive
chirality are left invariant after the action generated by \eqref{eq:killvect}. From the perspective
of $\ads{3}$, only one of the two inequivalent representations is not projected out by the quotient.
Thus, we conclude our discrete quotient preserves one half of the supersymmetry, a conclusion
that can also be reached by explicit computation.

\EPSFIGURE{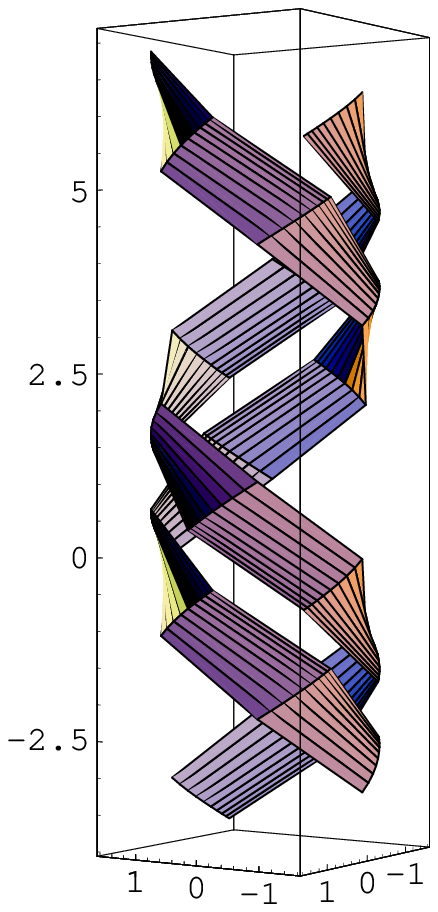}{Equal z slices.}

\subsection{Conformal boundary geometry}
\label{sec:boundary}

We seek the conformal boundary of our orbifold since we expect, a CFT living on it to give 
 the holographic description of string theory on our space.
A priori, there are two natural approaches to this problem, which are generically inequivalent. 
We could approach this in two ways: (a), we could determine the conformal 
boundary of the discrete quotient manifold \eqref{eq:fmetric}, or (b) we could study the quotient of the conformal boundary of  $\ads{3}$.    
As we will see that the discrete identification we are using induces a {\it conformal} transformation rather than an  isometry of the $\ads{3}$ boundary.  
Thus, approach (b) is not well defined here, but we will discuss the relevant transformations since they are useful later.


\subsubsection{Conformal boundary of the quotient}
\label{sec:confbound}

Global $\ads{3}$ has an $\bR\times S^1$ cylinder
as its conformal boundary,  located at $\rho\to\infty$ in the
coordinates \eqref{eq:global}.  The map to coordinates adapted to the orbifold \eqref{eq:adapted}  is:
\begin{equation}
  \begin{aligned}
    \cosh^2\rho &= \cosh^2z \, \cosh^2\phi + \sinh^2z \, \sinh^2\phi\,, \\
    \tan\tau &= \frac{\tan{t} - \tanh{z} \tanh\phi}{1 +
    \tanh{z} \tanh\phi \tan{t}}\,, \\
    \tan\theta &= \frac{\tanh\phi \tan{t} - \tanh{z}}{\tanh\phi +
    \tanh{z}\tan{t}}\,.
  \end{aligned}
 \label{eq:coordmap}
\end{equation}
We see that the $\ads{3}$ boundary at $\rho \rightarrow \infty$ can be reached by either taking $z \to \pm \infty$  or $\phi \to \pm \infty$ in the adapted coordinates.

To get some familiarity with the adapted coordinates, let us 
display global $\ads{3}$ as  a solid cylinder.
Then,  the equal $z$ and equal $t$ surfaces of
\eqref{eq:fmetric} in the adapted coordinates are displayed in Figs.~1 and~2.
An equal $z$ section is a helical strip (actually a cylinder since the opposite ends of the strip are identified) which winds up the global cylinder. The boundary of the quotient consists of two such strips on the boundary of the global cylinder, with each strip being at $z=\infty$ or $z=-\infty$. It is easy to see why the original $\ads{3}$ boundary splits into two pieces.  The action of the orbifold makes $\phi$ a compact variable. So the region between the two strips on the boundary of the global cylinder  is the locus of the points for which $\phi \rightarrow \pm \infty$. 



The metric on these conformal boundaries can be worked out using standard
techniques. Let us consider the coordinate transformation
\[
  \sinh z = \tan \theta \quad \quad \theta \in \left(-\frac{\pi}{2}\,,\frac{\pi}{2}\right)
\]
bringing infinity to a finite distance . The metric
of our discrete quotient \eqref{eq:fmetric} can be written as
\begin{equation*}
  g = \frac{l^2}{(\cos\theta)^2}\left((\cos\theta)^2 (-(dt)^2 +
  (d\phi)^2) + (d\theta)^2 + 4\sin\theta dt d\phi\right)~.
\end{equation*}
It is then clear that the metric on both conformal boundaries, located
at $\theta\to\pm \frac{\pi}{2}$ is given by
\begin{equation}
  \tilde{g} = \pm dt d\phi ~, 
 \label{eq:cbmetric}
\end{equation}
after a conformal rescaling.

The metric \eqref{eq:cbmetric} is locally that of flat two dimensional space in lightcone coordinates.
However, $\phi$ is an angular coordinate. Thus both conformal
boundaries have closed lightlike curves.  We will refer to such a geometry 
with a flat metric and one compact null direction as a {\it null cylinder}.
We explained that from the
$\ads{2}$ fibration perspective the space also has two boundaries, but that they are timelike lines.  The two points of view are reconciled by observing that  in $\ads{2}$ there is an electric field that dominates over the 
metric components
at large $z$, explaining the light-like nature of the 3d boundary.

\EPSFIGURE{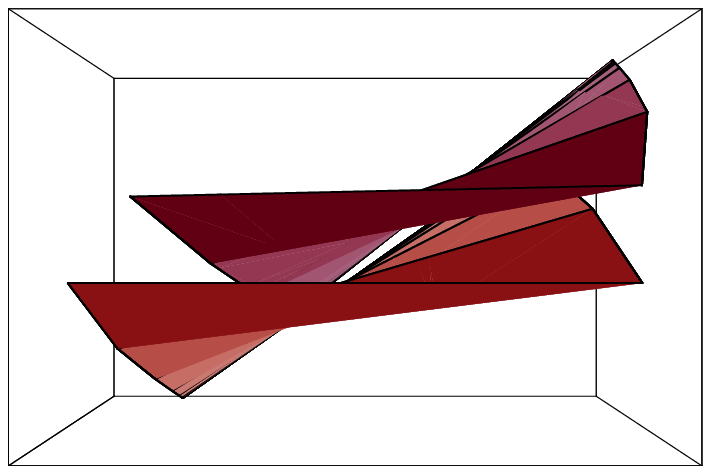}{Equal t slices: The equal $t$ sections, when displayed in
 global coordinates look like wedges that rotate in global AdS as $t$ 
 increases.\vspace{0.5cm}}

We can get some insight into such null cylinder geometries by examining the 
geometry of fixed $z$ surfaces in the metric (\ref{eq:fmetric}).  
The metric on any fixed $z$ surface is
\begin{equation}
  g = l^2(-dt^2 + d\phi^2 + 2\sinh(2z) \, dt d\phi)~,
\label{fixedzmet}
\end{equation}
where $\phi$ is  periodically identified. In fact this metric is 
conformally related to a boosted version of the metric on a time-like cylinder  swept out by $t$ and $\phi$ at $z=0$:
\begin{equation}
  g = l^2(-dt_0^2 + d\phi_0^2)~,
\label{cylinder}
\end{equation}
where $\phi_0$ is a circular direction with $\phi_0 \sim \phi_0 + 2\pi$.  To see this,  we boost \eqref{cylinder} with a rapidity $\eta$ 
to get the new coordinates
\begin{equation}
  \tilde{t} =  t _0\, \cosh(\eta) -  \phi_0 \, \sinh(\eta) ~~~;~~~
  \tilde{\phi} = -t_0 \, \sinh(\eta) +  \phi_0 \, \cosh(\eta) \, .
\end{equation}
These new coordinates for the cylinder are identified as
$(\tilde{t},\,\tilde{\phi}) \sim (\tilde{t},\,\tilde{\phi}) + 
2\pi\,(-\sinh\eta,\,\cosh\eta)$. It is convenient to choose an adapted
coordinate system in which the identification acts at a fixed time
\begin{equation}
  \tau \equiv \tilde{t} \, \cosh\eta + \tilde{\phi} \, \sinh\eta = t_0 ~~~;~~~
  \beta \equiv   - \tilde{t} \, \sinh\eta + \tilde{\phi} \cosh{\eta}
\end{equation}
in terms of which the identifications are $\beta \sim \beta + 2\pi 
\cosh(2\eta)$ at fixed $\tau$.  This leads to a metric
\begin{equation}
  g =  {1 \over \cosh^2(2\eta)}  [ -d\tau^2 + d\beta^2 + 2\sinh(2\eta) \, d\tau \, d\beta]
\end{equation}

This is conformal to the metric on fixed $z$ slices (\ref{fixedzmet}).   A 
further rescaling of coordinates
\begin{equation}
  \tau \equiv \cosh(2\eta) \, t ~~~;~~~ \beta \equiv \cosh(2\eta) \, \phi
\end{equation}
leads to the metric on the metric on the boosted cylinder
\begin{equation}
  g = -dt^2 + d\phi^2 + 2\sinh(2\eta) \, dt \, d\phi
\end{equation}
exactly producing the metric (\ref{fixedzmet}) on the fixed $z$ slices of our 
orbifold geometry.  The radial coordinate $z$ plays the role of the rapidity 
parameter in the boost.  As $|z| \to \infty$ these metrics 
systematically approach the ``null cylinder" of the conformal boundary.

\subsubsection{Quotient of the conformal boundary}

Alternatively, we can study how the generator of the discrete
identification \eqref{eq:killvect} acts on the original conformal boundary $\bR\times S^1$ of $\ads{3}$:
\[
  g = -(d\tau)^2 + (d\theta)^2 \,.
\]
The isometry group $\SO(2,2)$ in the bulk becomes the conformal
group of the boundary metric. Its generators become conformal Killing vectors.
One can write these generators explicitly, by rewriting $J_{ab}$ in terms
of the global coordinates \eqref{eq:global} and evaluating them on the
boundary: 
\begin{equation}
  \begin{aligned}
    J_{01} & = \partial_\tau \\
    J_{02}  &= -(\cos\theta\sin\tau\partial_\tau + \cos\tau\sin\theta
    \partial_\theta) \\
    J_{03}  &= -\sin\theta\sin\tau\partial_\tau + \cos\tau\cos\theta
    \partial_\theta \\
    J_{12}  &= \cos\theta\cos\tau\partial_\tau - \sin\tau\sin\theta
    \partial_\theta  \\
    J_{13} & = \sin\theta\cos\tau\partial_\tau + \sin\tau\cos\theta
    \partial_\theta \\
    J_{23}  &= -\partial_\theta \,.
  \end{aligned}
 \label{eq:confkill}
\end{equation}
Therefore, the maximal compact subgroup of $\SO(2,2)$ that is 
generated by $\{J_{01}\,,J_{23}\}$ is an isometry of the boundary metric, 
whereas the remaining generators are conformal Killing vectors. Indeed, 
\begin{equation}
  \left(\CL_{J_{0k}+iJ_{1k}} g\right)_{ab} = 2 \Omega_k g_{ab} \,,
\end{equation}
where $g_{ab}$ stands for the boundary metric and $(\Omega_2, \Omega_3) = e^{i\tau}
(\cos\theta\,,\sin\theta)$.

This analysis shows that the generator of our orbifold \eqref{eq:killvect} is not
an isometry of the boundary metric, but it is a conformal Killing vector.  Thus it is not clear what a discrete quotient of the $\ads{3}$ boundary by this generator should mean.
For later reference, the explicit expression 
for the generator of this discrete action on  $\bR\times S^1$ is
\begin{equation}
  \xi_{\rho\to\infty} = -\frac{1}{2}\sin (\tau-\theta) \left(\partial_\tau -
  \partial_\theta \right) \,.
 \label{eq:bounkillvect}
\end{equation}
Some comments are in order:
\begin{itemize}
\item
  If we forget about the factor $\sin (\tau-\theta)$, the identification
  would certainly generate a closed lightlike direction, matching our
  previous conclusion.
\item
  The action on the conformal boundary $\bR\times S^1$ has fixed points
  located at $\tau = \theta (\text{mod}\,\pi)$. This may seem surprising because
  our bulk analysis allowed us to write the generator as 
  $\partial/\partial\phi$ everywhere, both in the bulk and the boundary,
  which shows there are no fixed points. However, this  misses the fixed point 
  at $\phi \to \pm \infty$. From (\ref{eq:coordmap}), it is clear that $\phi \to \pm\infty$ 
  corresponds to $\tau = \theta$. 
\end{itemize}

In summary, we have established that the conformal boundary of our orbifold consists of two null cylinders.  In Sec.~\ref{sec:conforb} we will explore the possibility of directly orbifolding the dual to $\ads{3}$ by the above discrete conformal transformation
to generate the dual to our orbifold.

\subsection{Geodesics}
\label{sec:geodesics}

Geodesics extremize the length of curves connecting two points. For a smooth 
curve $x^\mu(\lambda)$, the length $S$ is 
\begin{equation}
  S=\int d \lambda  \sqrt{|g_{\mu \nu}{d x^\mu \over d \lambda} 
  {d x^\nu \over d \lambda}|}\,.
\end{equation}
The problem is identical to the extremization of the 
action in Lagrangian particle mechanics. In fact, with affine parameterization, the geodesic equations can be obtained by variation of a simpler
Lagrangian
\begin{equation}
  L=g_{\mu \nu} {dx^\mu \over d\sigma}{dx^\nu \over d \sigma}\,,
\end{equation}
where $\sigma$ is now the affine parameter. For the metric in 
\eqref{eq:fmetric}, this yields (we have set 
$l=1$)
\begin{equation}
  L=-\dot{t}^2+\dot{\phi}^2+\dot{z}^2+2\sinh(2z) \dot{\phi}\dot{t}\,,
\end{equation}
where the dots stand for derivatives with respect to the affine
parameter. Since the Lagrangian is independent of both $t$ and $\phi$, we have
the conserved momenta:
\begin{equation}
  \begin{aligned}
    P_t &= -2\dot{t} + 2\sinh(2z)\dot{\phi} = E\,, \\
    P_\phi & =  2 \dot{\phi} + 2\sinh(2z) \dot{t} = m\,, \\
  \end{aligned}
\end{equation}
which imply
\begin{equation}
  \begin{aligned}
    \dot{t} &= \frac{m \sinh(2z)-E}{2\cosh^2(2z)}~, \\
    \dot{\phi} & =  \frac{E \sinh(2z)+m}{2 \cosh^2(2z)}~.
  \end{aligned}
\end{equation}
In addition, the Hamiltonian $H$  is equal to the Lagrangian $L$ in our 
case and therefore the Hamiltonian constraint is
\begin{equation}
  -\dot{t}^2+\dot{\phi}^2+\dot{z}^2+2\sinh(2z) \dot{\phi}\dot{t}=k
\end{equation}
where we can choose the affine parameterization such that $k=-1,0,1$ for 
time-like, null and space-like geodesics respectively. From this we obtain
\begin{equation}
  \dot{z}=\sqrt{k+{E^2-m^2 \over 4 \cosh^2(2z)}-Em{\sinh(2z) \over 2\cosh^2 
  (2z)}}
 \label{eq:zdot}
\end{equation}
For generic values of the constants of motion $E$ and $m$, this equation admits the following solutions:
\begin{eqnarray}
  k=0 & \quad & \sinh(2z)=\frac{E^2-m^2}{2mE}-
  \frac{(\sigma_0 -E\cdot m\sigma)^2}{2Em} \nonumber \\
  k=-1 & \quad & \sinh(2z) =-\frac{E\cdot m}{4} + \frac{1}{2}
  \cos(\sigma_0-2\sigma)\sqrt{E^2-m^2-4+ (E\cdot m/2)^2} \nonumber  \\
  k=1 & \quad & \sinh(2z)= \frac{1}{2}\sinh (2\sigma + \sigma_0)
  \sqrt{E^2 - m^2 + 4 - (E\cdot m/2)^2} +
  \frac{E\cdot m}{4}
 \label{eq:gengeo}
\end{eqnarray}

Whenever $E=0$, $m=0$ or both, some of the previous solutions
become singular, and so these cases have to be dealt with separately.  Notice that
from the square of equation \eqref{eq:zdot}, whenever $E=m=0$, timelike geodesics $(k=-1)$ do not
exist, and we are  left with the trivial solution $z=z_0$, $t=t_0$ and $\phi=\phi_0$,
or the spacelike geodesic $(k=1)$, $z=\pm\sigma + z_0$, $t=t_0$ and $\phi=\phi_0$.
If we keep $E=0$, but consider non-vanishing angular momentum, the solution can be obtained
from the appropriate limit in \eqref{eq:gengeo}.  Such geodesics are space-like.
Actually, if $m$ is an integer, they only make sense for $m=0,1,2$. For example,  for $m=2$ the solution is given by
\begin{equation}
  \sinh 2z = Ce^{2\eta\sigma} \quad , \quad \eta=\pm \,.
\end{equation}

If the angular momentum vanishes $(m=0)$, but we consider non-vanishing energy, one can find
the lightlike geodesic
\begin{equation}
  \sinh\,2z = \eta\,E\,\sigma + \sigma_0~.
 \label{eq:singgeo}
\end{equation}
The spacelike solutions can be obtained from the corresponding limit in \eqref{eq:gengeo}.
Among the timelike ones, if $\dot{z}\neq 0$, the solutions only exist for $|E|>2$. If $z=z_0$,
then we can allow $|E|\geq 2$, and the solutions are given by $t=-\frac{2}{E}\sigma + t_0$
and $\phi = \sqrt{1-4/E^2}\sigma + \phi_0$.
Note that for $E^2=m^2$, one can find lightlike geodesics consisting of constant
$z=z_0$, and linear dependences in the affine parameter both for $t$ and $\phi$.
Finally, there are timelike geodesics at any constant $z$ and $\phi$ (this is easily seen by solving $\dot{\phi} = \dot{z} = 0$).   These geodesics descend from trajectories in global AdS that spiral around the origin. 
The spiral is ``unwound'' by the twisted coordinate system  (\ref{eq:adapted}) that we have chosen from the perspective of the embedding in global AdS (see Fig.~2).

It is interesting to study which geodesics causally connect the two boundaries at $z=\pm \infty$. 
The lightlike geodesics $(k=0)$ written in \eqref{eq:gengeo},
do not connect the two boundaries. Indeed, these are parabolas,
so the best that we can do is to arrange initial conditions such that
the geodesic starts at one boundary and gets as close as we want to the second conformal boundary.
On the other hand, the lightlike geodesics with  vanishing angular momentum  in 
\eqref{eq:singgeo} provide a causal connection between the two boundaries.  We can find the orbit
in the $\{t,\,z\}$ plane in this case, and compute the coordinate time required for the
null particles to go from $z\to -\infty$ to the other boundary at $z\to \infty$. The result is given by
the finite expression
\begin{equation}
  t(z=+\infty)-t(z=-\infty) = -\eta\frac{\pi}{2} \,.
\end{equation}
Timelike geodesics show an oscillatory behavior and they do not connect both boundaries.
In sum, the only way to connect both boundaries with a geodesic is to use the null trajectory described above, or to consider unphysical extreme initial quantum numbers such as $E\to\infty$
and $m=\text{constant}$.

\subsection{Excursions in the complex coordinate plane}
\label{excursions}

While discussing holography for our orbifold, we will see that the symmetries and structure of the complexified manifold play an important role.  First of all, notice that while our metric 
\begin{equation}
ds^2 = -dt^2 + dz^2 + d\phi^2 + 2 \, \sinh(2z) \, dt \, d\phi
\label{metricagain}
\end{equation}
is non-singular everywhere for real $\phi$, $z$ and $t$, the determinant
\begin{equation}
\det{g} = - \cosh^2(2z)
\end{equation}
vanishes at $z = i\pi/4$.    It is not apparent at this point that  this singularity in the complex coordinate plane should play any role in the physics,  but as we will see later, this singularity make a crucial appearance in the analysis of holography. 

For later purposes it will be very useful to understand the symmetries of (\ref{metricagain}) as a metric on the space of complexified coordinates.

\paragraph{Discrete symmetries in the complex coordinate plane: }

In the global parametrization (\ref{eq:global}), the following discrete complex transformations of coordinates leave invariant points  of the  quadric defining $\ads{3}$ in $\bR^{2,2}$ (\ref{embedding}):
\begin{equation}
  \rho \to \rho + i\,\pi \quad , \quad \tau \to \tau + \pi \quad , \quad
  \theta\to \theta + \pi~,
\end{equation}
In the coordinate system adapted to our orbifold (\ref{eq:adapted}), there is a richer structure due to the combination of trigonometric and hyperbolic functions in the parametrization.
Three discrete complex coordinate transformations that leave the embedding invariant are
\begin{eqnarray}
  z &\to& z \quad , \quad t \to t + \pi \quad , \quad
  \phi \to \phi + i\,\pi~,
 \label{eq:disa} 
\\
  z &\to& z + i\,\pi \quad , \quad t \to t + \pi \quad , \quad
  \phi \to \phi~,
 \label{eq:disb}
\\
  z &\to& z +i\,\pi \quad , \quad t \to t \quad , \quad
  \phi \to \phi + i\,\pi~.
 \label{eq:disc}
\end{eqnarray}
These complex transformations of coordinates map real physical points on the orbifold to themselves. Later we will see that the solutions to the wave equation on our orbifold will reflect this symmetry in complex coordinate plane.


\paragraph{Motions through the complex coordinate plane:  } 

In addition, certain discrete complex coordinate transformations  generate transformations between real physical points in the orbifold.  First, we show a transformation connecting the two components of the conformal boundary of the spacetime.   Consider (\ref{eq:coordmap}) which relates the global and adapted coordinates.  We are interested in boundary points, namely $z\to\pm \infty$.   In global coordinates this amounts to  $\rho \to \infty$ in the first of the 
equations (\ref{eq:coordmap}).  The second and third equations can then be rewritten in terms 
of a sign $\eta$ corresponding to the boundary at $z \rightarrow \eta \infty$ :
\begin{equation}
  \begin{aligned}
    \tan\tau &= \frac{\tan{t} - \eta \tanh\phi}{1 +
    \eta \tanh\phi \tan{t}}\,, \\
    \tan\theta &= \frac{\tanh\phi \tan{t} - \eta}{\tanh\phi +
    \eta \tan{t}}\,.
  \end{aligned}
 \label{eq:infcoordmap}
\end{equation}
Now, consider the discrete transformation 
\begin{equation}
  t= t^\prime + \frac{\pi}{2} \quad , \quad \phi = \phi^\prime + i\,\frac{\pi}{2}~,
 \label{eq:infz}
\end{equation}
defined  on the boundary $z\to\infty$.  
Taking the expressions \eqref{eq:infcoordmap} with $\eta=+1$, and using the transformations \eqref{eq:infz}, we  find
the description of a point in global $\ads{3}$ coordinates sitting at $\rho\to\infty$ with $(\tau^\prime,\,\theta^\prime)$ given
by
\begin{equation}
  \begin{aligned}
    \tan\tau^\prime &= \frac{\tan{t^\prime} + \tanh\phi^\prime}{1 -
    \tanh\phi^\prime \tan{t^\prime}}\,, \\
    \tan\theta^\prime &= \frac{\tanh\phi^\prime \tan{t^\prime} + 1}{\tanh\phi^\prime -
    \tan{t^\prime}}\,.
  \end{aligned}
\end{equation}
But this is precisely the description of a point in the second boundary $z\to -\infty$ obtained
by evaluating \eqref{eq:infcoordmap} with $\eta=-1$.  Therefore the complex transformations \eqref{eq:infz} map points
in the boundary $z\to\infty$ to points in the second boundary at $z\to -\infty$.

In fact, (\ref{eq:infz}) can be extended to a complex transformation mapping real points on the orbifold for all $z>0$ to corresponding points at $z<0$.  Keeping in mind the relation
\begin{equation}
  \cosh^2\rho = \cosh^2 z\,\cosh^2\phi + \sinh^2 z\,\sinh^2\phi~.
\end{equation}
let us extend the transformation \eqref{eq:infz} by the following action on $z$
\begin{equation}
  z = z^\prime + i\,\frac{\pi}{2}~.
\end{equation}
This transformation preserves $\cosh^2\rho = \cosh^2\rho^\prime$. Since $\tanh\,z= -(1/\tanh\,z^\prime)$, the point
in global $\ads{3}$ described by $(t^\prime,\,\phi^\prime,\,z^\prime)$ is given by
\begin{equation}
 \begin{aligned}
    \tan\tau^\prime &= \frac{\tanh z^\prime\tan{t^\prime} + \tanh\phi^\prime}{1 -
    \tanh z^\prime \tanh\phi^\prime \tan{t^\prime}}\,, \\
    \tan\theta^\prime &= \frac{\tanh\phi^\prime \tan{t^\prime} + \tanh z^\prime}{\tanh\phi^\prime -
    \tanh z^\prime\tan{t^\prime}}\,.
  \end{aligned}
\end{equation} 
Thus, we can identify $(t^\prime,\,\phi^\prime,\,z^\prime)$ with $(t,\,\phi,\,-z)$ by explicitly evaluating (\ref{eq:coordmap}).

We will see that these discrete transformations in the complex coordinate plane that map real physical points to each other have an intimate relation to the structure of holographic duality in the orbifold.

\section{Scalar field theory}
\label{sec:modesolns}

Since the AdS orbifold constructed in the previous section has two boundaries 
we expect that the dual field theory will take a novel form.  It could be a 
product of two independent CFTs, or the two components could be identified or 
entangled.   The first step in understanding both bulk and boundary dynamics is to 
study scalar field theory in the background (\ref{eq:fmetric}) with the orbifold identification $\phi \sim \phi  + 2\pi$.
In Lorentzian AdS spaces we expect a spectrum of normalizable modes 
corresponding to states in the dual CFT and non-normalizable modes 
corresponding to sources \cite{BKL}.   We will find a basis of such mode 
solutions and use them to infer facts about the structure of the CFT.

The wave equation for a massive scalar is
\begin{equation}
  (\Box-\mu^2) \Psi =0
\end{equation}
where $\mu$ includes the effects of a curvature coupling, if there is one, since 
the curvature is constant in our orbifold.  For the metric \eqref{eq:fmetric}, 
this becomes,
\begin{equation}
  \left\{-\frac{\partial^2}{\partial t^2}+ \cosh^2(2z) \, 
  \frac{\partial^2}{\partial z^2} + \frac{\partial^2}{\partial \phi^2} 
  + 2\sinh(2z) \,\frac{\partial^2}{\partial t \partial \phi} + \sinh(4z)\,
  \frac{\partial}{\partial z}-\mu^2 \cosh^2(2z)\right\} \Psi =0
\label{waveeqn}
\end{equation}

\subsection{Unitary representations of $SL(2,R)$ and fluctuating states}
\label{sec:representations}

Fluctuating states on the orbifold spacetime should lie in a unitary representation of the isometry group $\SL(2,\bR) \times \U(1)$.   
Following \cite{BKL} and references therein we should look for a highest weight representation of $\SL(2,\bR)$.
To this end, define a new basis for the  $\fsl(2,\bR)$ generators in 
(\ref{isometries}) as 
\begin{eqnarray}
  \CL_0 &=& i \chi_1 = {i \over 2} {\partial \over \partial t} \nonumber  \\
  \CL_+ &=& \chi_2 - i \chi_3 = \frac{e^{2it}}{2}\left[\tanh(2z) 
  {\partial \over \partial t} + {1\over \cosh(2 z)} 
  {\partial \over \partial \phi} - i 
  {\partial \over \partial z}\right]  \label{sl2gens} \\
  \CL_- &=& -(\chi_2 + i \chi_3) = -\frac{e^{-i2t}}{2}\left[ \tanh(2z) 
  {\partial \over \partial t} + {1\over\cosh(2 z)} 
  {\partial \over \partial \phi} + i 
  {\partial \over \partial z} \right] \, , \nonumber 
\end{eqnarray}
in terms of which the $\fsl(2,\bR)$ algebra is $[\CL_0, \CL_\pm] = \mp \CL_\pm$
and $[\CL_+,\CL_-] = 2 \CL_0$.   Highest weight states satisfy 
$\CL_+  |h\rangle = 0$ and $\CL_0 |h\rangle = h |h \rangle$.  A complete 
highest weight representation is constructed as $\CL_-^n | h \rangle$ for 
$n \geq 0$.    In such a highest weight representation the $\fsl(2,\bR)$ 
Casimir is given by
\begin{equation}
  2\CL^2 = 2\CL_0^2 - (\CL_+ \CL_- + \CL_- \CL_+) = 2h(h-1)
 \label{casimir}
\end{equation}
In terms of the explicit differential operators in (\ref{sl2gens}), it is easy to 
show that the Casimir equation can be written as
\begin{equation}
  4 \CL^2 \psi = \Box \psi = 4h(h-1) \psi
\end{equation}
so  a solution to the wave equation (\ref{waveeqn}) that is in a highest 
weight representation of the isometries satisfies
\begin{equation}
  \mu^2 = 4h(h-1)  ~~~~\Longrightarrow~~~~ h_\pm ={1\over 2} \pm {1\over 2} 
  \sqrt{1 + \mu^2}
\end{equation}
Highest weight states solve the equation
\begin{equation}
  \CL_+ \Psi(t, z, \phi) \equiv \CL_+ \psi(t,z) e^{im\phi} = 0 ~~~;~~~ 
  \CL_0 \Psi(t, z, \phi) \equiv \CL_0 \psi(t,z) e^{im\phi} = h \psi(t,z) 
  e^{i m \phi}
\end{equation}
where we have have Fourier transformed in  $\phi$ since 
$\xi = (i/2) (\partial/\partial \phi)$ is an isometry.  The second equation 
yields $\Psi = \tilde{\psi}(z) e^{-2iht} e^{i m \phi}$ and the 
remaining first order differential equation is easily solved to give
\begin{equation}
  \Psi(t,z,\phi) = C \, [\cosh(2z)]^{-h} \, e^{m \tan^{-1}[\tanh(z)]} \, 
  e^{-2iht} \, e^{i m\phi}
 \label{highestwave}
\end{equation}
where C is a normalization constant.

We first examine the behavior of (\ref{highestwave}) at the 
orbifold boundaries $z \rightarrow \pm \infty$.   In this limit 
$\tan^{-1}[\tanh(z)] \rightarrow \pm \pi/4$, and so  that as 
$z \rightarrow \pm \infty$, $| \Psi | \rightarrow 0$ if $h>0$ and 
$| \Psi | \rightarrow \infty$ if $h < 0$.   For $\mu^2 > 0$, normalizable 
modes arise from $h_+$ and non-normalizable modes from $h_-$.  When $m = 0$ 
the normalizable highest weight state is a lump localized around $z = 0$.   
When $m < 0$ ($m>0$) the state is a lump is localized at some $z_0 <0$ 
($z_0 > 0$) (see Fig.~3).    The non-normalizable modes diverge when 
$z \rightarrow \pm \infty$, but they too are asymmetric when $m \neq 0$.   
For negative (positive) $m$, these modes grown much faster in the region 
$z < 0$ ($z > 0$). This behavior has the  consequence that the physics of modes with 
positive $m$ is largely concentrated at positive $z$ while 
the physics of negative $m$ is largely concentrated at negative 
$z$.




A complete highest weight representation is obtained by 
acting on \eqref{highestwave} with the operator $\CL_-^n$.  The descendants are characterized conveniently in terms of 
\[
  \Psi_h (t\,,\phi\,,z) = e^{-2iht}\,e^{im\phi}\,\Psi_h(z)\equiv
  e^{-2iht}\,e^{im\phi}\,\Bigl(\cosh 2z\Bigr)^{-h}\,e^{m\tan^{-1}[\tanh (z)]}
  \quad \forall h ~.
\]
Since $\Psi^\prime_h (z) = \Psi_{h+1}(z)\left(-2h\sinh 2z + m\right)$,
\begin{equation}
  \CL_-\Psi_h (t\,,\phi\,,z) = i\left(2h\sinh 2z - m\right)
  \Psi_{h+1} (t\,,\phi\,,z)\quad \forall h~.
 \label{eq:desc}
\end{equation}
In general,  $\CL_-^n\Psi_h (t\,,\phi\,,z)= g(n,z) \Psi_{h+n}(t\,,\phi\,,z)$.  We will derive $g(n,z)$ in the next subsection.

\subsection{Normalizable vs. non-normalizable modes and analytic structure}
\label{sec:modes}

\EPSFIGURE{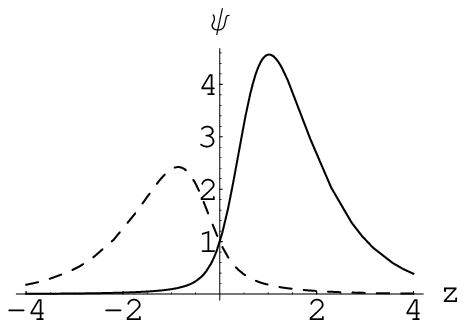}{Mode solutions: The dashed solution corresponds to $m<0$, whereas the solid one corresponds to $m>0$.}

The non-normalizable modes which are necessary for the AdS/CFT correspondence are expected to be in a mixture of $SL(2,R)$ representations \cite{BKL} and so it is more convenient to directly solve for them from the wave equation \eqref{waveeqn}.   On the way this will also give us the function $g(n,z)$ for the descendant normalizable states.

We start with the ansatz
\begin{equation}
  \Psi(t,z,\phi)=e^{-i \omega t + i m \phi} \psi(z)~.
\label{psi1}
\end{equation}
Inserting this into \eqref{waveeqn}, one obtains:
\begin{equation}
  \cosh^2(2z) \frac{d^2 \psi}{dz^2}+ \sinh(4z)\frac{d\psi}{dz}+
  \Bigl(\omega^2-m^2 + 2\omega m \sinh(2z) -\mu^2 \cosh^2(2z)\Bigr)\psi=0~.
\end{equation}
Consider the change of variables
\begin{eqnarray}
  \psi (z) &= & \Bigl(\cosh 2z\Bigr)^b\,e^{a \tan^{-1}[\tanh(z)]}\,\chi (z) \label{ansatz}\\
  y &=& \frac{1}{2}(1 + i \sinh 2z)\,,
\end{eqnarray}  
The function $\chi(z)$ then satisfies the hypergeometric equation 
\begin{equation}
  y(1-y) \frac{d^2 \chi}{dy^2} + \Bigl(C-(1+A+B)y\Bigr) \frac{d\chi}{dy} 
  - AB \chi =0
 \label{eq:hyper}
\end{equation}
provided 
\[
4b^2-\omega^2+m^2-a^2+(4ba + 2 \omega m)i=0.
\]
This equation is solved by $b= \mp\frac{\omega}{2}$ and $a= \pm m$ -- 
either choice of signs will give identical solutions.  So from now on we choose the upper signs.
The constant parameters $A,B$ and $C$ appearing in \eqref{eq:hyper} are determined
in terms of $\{m\,,\omega\,,\mu^2\}$ by:
\begin{equation}
  \begin{aligned}
    A &=  -\frac{\omega}{2}+\frac{1}{2} - \frac{1}{2}\sqrt{1+\mu^2} =  -{\omega \over 2} +h_{-}\\
    B &= -\frac{\omega}{2} + \frac{1}{2} + \frac{1}{2}\sqrt{1+\mu^2}= -{\omega \over 2} +h_{+} \\
    C &= 1 -  \Bigl( \frac{\omega + im}{2}\Bigr),
  \end{aligned}
\label{case1}
\end{equation}    

In order to determine the normalizable and non-normalizable solutions of 
\eqref{eq:hyper}, we need to determine the asymptotic behavior of the solutions
as $z \rightarrow \pm \infty$. For this reason, it is most convenient to write 
the solution in terms of hypergeometric functions with arguments $1/y$:
\begin{equation}
  \chi(z) = c_1 \, y^{-A}{}_2F_1(A,A-C+1;A-B+1; {1 \over y})
  +c_2 \, y^{-B}{}_2F_1(B,B-C+1;B-A+1,{1 \over y})
 \label{eq:gensoln}
\end{equation}
The solution is written as a sum of two independent solutions of the 
hypergeometric equation. This is a convenient basis of solutions to isolate 
the normalizable and non-normalizable solutions. 

\paragraph{Asymptotics}
As $z\rightarrow \infty$,  ${1 \over y} \rightarrow 0$, and the hypergeometric 
functions in \eqref{eq:gensoln} approach 1.
Thus
\begin{equation}
  \chi \rightarrow c_1 \, y^{-A}+ c_2 \,  y^{-B}
\end{equation}
For the full  solution $\Psi(t,\phi,z)$ in \eqref{psi1}, using \eqref{ansatz}, this 
implies
\begin{equation}
  \Psi(t,\phi,z) \rightarrow e^{-i\omega t + im\phi}(c_1 \, (e^{2z})^{-h_-}
  +c_2 \,  (e^{2z})^{-h_+})
\end{equation}
As $z\rightarrow -\infty$, the asymptotic behavior is 
\begin{equation}
  \Psi(t,\phi,z)  \rightarrow 
e^{-i \omega t+ i m\phi}(c_1 \,  (e^{-2z})^{-h_-}
  +c_2 \, (e^{-2z})^{-h_+})
\end{equation}
Hence the solution with coefficient $c_2$ decays at both boundaries while the solutions multiplying $c_1$ grows at both boundaries and is non-normalizable.

\subsubsection{Normalizable solutions}
\label{sec:normalizable}

We seek a basis for the fluctuating normalizable mode solutions for the scalar field.   Candidates are given by the modes that decay at infinity:
\begin{equation}
  \Psi_{\rm norm}(t,\phi,z)=e^{-i\omega t + im \phi}e^{m \tan^{-1}(\tanh(z))}
  (\cosh(2z))^{-{\omega \over 2}} ~y^{({{\omega \over 2}-h_{+}})}~
  {}_2F_1(-{\omega \over 2}+h_{+}, h_{+} + i{m \over 2}; 2h_{+};{1 \over y})
 \label{eq:solnorm}
\end{equation}
We have already displayed a suitable basis of modes via the analysis of unitary SL(2,R) representations given above.  However we will see that the same quantized spectrum of solutions arises from (\ref{eq:solnorm}) by requiring that 
the mode solutions should be single valued in the complex $z$ plane.   Presumably this requirement is necessary for unitarity.   

To this end, let us study the behavior of (\ref{eq:solnorm}) as $y \rightarrow 0$ or, equivalently, 
as $z \rightarrow i\pi/4$.
In Sec.~\ref{excursions} we noted that $z = i\pi/4$ is a location in the complex 
coordinate plane at which the determinant of the metric (\ref{eq:fmetric}) vanishes; namely this is a singular point in the complex coordinate plane.   Using the transformation formula
\begin{eqnarray*}
  {}_2F_1(\ta,\tb;\tc;{1 \over y})&=&{\Gamma(\tc) \Gamma(\tb-\ta) \over 
  \Gamma(\tb)\Gamma(\tc-\ta)}(-{1 \over y})^{-\ta}
  {}_2F_1(\ta, 1-\tc+\ta;1-\tb+\ta;y) \\
  &&~~~~~~~+{\Gamma(\tc) \Gamma(\ta-\tb) \over \Gamma(\ta)\Gamma(\tc-\tb)}
  (-{1 \over y})^{-\tb}{}_2F_1(\tb,1-\tc+\tb;1-\ta+\tb;y)
\label{transformation}
\end{eqnarray*}
we find
\begin{multline}
  \Psi~~ \sim (-1)^{(\omega/2 - h_+)}
  {\Gamma(2h_{+}) \Gamma({\omega + im \over 2}) \over \Gamma(h_+ + {i m \over 2})
  \Gamma(h_++{\omega \over 2})} {}_2F_1(-{\omega \over 2} + h_+, 
  -{\omega \over 2} + h_{-}; 1 - {{\omega + im} \over 2}; y) \\
  + (-1)^{-(h_+ + im/2)}{\Gamma(2h_+) \Gamma(-{\omega + im \over 2}) \over 
  \Gamma(-{\omega \over 2}+h_+)\Gamma(h_+ - i {m \over 2})}~
  y^{({\omega + im \over 2})}{}_2F_1(h_+ + i{m\over 2}, h_{-}+i{m\over 2};
  1+i{m\over 2}+{\omega\over 2}; y) ~.
 \label{eq:z0}
\end{multline}
Thus, the second term in \eqref{eq:z0} gives rise to a multivalued function in the complex $z$-plane.\footnote{The
nature of the cut is somewhat different when $m=0$ as opposed to $m\neq 0$.}
Requiring a single valued scalar wave function
forces us to require the vanishing of the second coefficient in  \eqref{eq:z0}.
This is achieved by 
\begin{equation}
  -{\omega \over 2}+h_+=-n~~~~~~n \in \bZ^+ \cup \{0\}~,
 \label{eq:qcond1}
\end{equation}
which ensures that $\Gamma(-{\omega \over 2}+h_+)$ has a vanishing or negative 
integer argument which makes it diverge.  Thus, by looking at (\ref{eq:solnorm}) and its complex conjugate, 
we find that the allowed states have energies (eigenvalues of $i\partial/\partial t$) 
\begin{equation}
  E_n =   \pm 2(h_+ + n)~~~~~~n \in \bZ^+\cup \{0\}~.
 \label{eq:qcond2}
\end{equation}
This quantization condition was also obtained from the 
analysis of highest weight representations of the isometry group in the 
previous subsection. Notice that the spectrum is entirely independent of the angular momentum quantum number $m$.      It is intriguing that requiring regularity of the solutions at a point the complex $z$ plane where the complexified metric degenerates has correctly selected the quantized spectrum of modes expected from SL(2,R) representation theory.

With this quantization the positive energy solutions in 
\eqref{eq:solnorm} become
\begin{eqnarray}
  \Psi&=&e^{-i(2h_++2n) t + im \phi}e^{m \tan^{-1}(\tanh(z))}
  (\cosh(2z))^{-(h_++n)} ~\Bigl({1 \over 2}(1+i\sinh(2z))\Bigr)^n~ \nonumber \\
  && ~~~~~~~\times{}_2F_1(-n, h_{+} + i{m \over 2}; 2h_{+};
  {\Bigl({1 \over 2}(1+i\sinh(2z))\Bigr)}^{-1})
 \label{eq:normal}
\end{eqnarray}
We can write the hypergeometric function appearing in this equation in terms of
Jacobi polynomials by using 
\begin{equation}
  {}_2F_1(-n, \alpha+\beta +n+1; \alpha+1;x)={n! \Gamma(2h_+) \over 
  \Gamma(2h_++n)}P_n^{(\alpha,\beta)}(1-2x)
\end{equation}
where $P_n^{(\alpha,\beta)}$ are the Jacobi polynomials defined as:
\begin{equation}
  P_n^{(\alpha,\beta)}(u)={\Gamma(\alpha+n+1) \over n!\Gamma(\alpha+\beta+n+1)}
  \sum_{k=0}^n {n \choose k} 
  {\Gamma(\alpha+\beta+k+n+1) \over 2^k \Gamma(\alpha+k+1)}(u-1)^k ~.
\end{equation}
Using $\alpha=2h_+-1$ and $\beta={-h_+-n + i {m \over 2}}$, \eqref{eq:normal} 
becomes
\begin{eqnarray}
  \Psi_n&=&e^{-i(2h_++2n) t + im \phi}e^{m \tan^{-1}(\tanh(z))}
  (\cosh(2z))^{-(h_++n)}  \nonumber \\
  && ~~~~\times {\Gamma(2h_{+}) \over \Gamma(h_{+} + i{m \over 2})}
  \sum_{k=0}^n {n \choose k} {\Gamma(h_{+}+ i{m \over 2} +k) 
  \over \Gamma(2h_{+}+k)} (-1)^k\Bigl({1 \over 2}(1+i\sinh(2z))\Bigr)^{n-k}
\end{eqnarray}
Earlier we derived the fluctuating states by analyzing a highest weight representation of the isometries.   The above
analysis completes this derivation by giving an explicit form for the states as can checked from the 
equation $\Psi_n={\cal L}_-^n \Psi_h$ where $\Psi_h$ was defined in \eqref{highestwave}.   Thus we see that requiring analyticity of the 
scalar wavefunctions in the complex coordinate plane is equivalent here to requiring that they transform in a unitary representation.

\subsubsection{Non-normalizable solution}
\label{sec:nonnorma}

The non-normalizable part of the solution \eqref{eq:gensoln} is
\begin{equation}
  \Psi_{\rm non-norm}=e^{-i\omega t +im \phi}e^{m \tan^{-1}(\tanh(z))}
  (\cosh(2z))^{-{\omega \over 2}} ~y^{({{\omega \over 2}-h_{-}})}~
  {}_2F_1(-{\omega \over 2}+h_{-}, h_{-} + i{m \over 2}; 2h_{-};{1 \over y})
 \label{eq:solnotnorm}
\end{equation}
The non-normalizable modes in AdS are related to sources in the dual field theory and hence should not be quantized \cite{BKL}.   
These modes can also be thought of as a momentum space representation of the bulk-boundary propagator.   Notice that the non-normalizable character of a mode is not changed by adding any normalizable modes to it.   This ambiguity in the selection of a basis of non-normalizable modes is related to the selection of a vacuum state for the spacetime and for the dual field theory \cite{BKL,BKLT,tommy}. 
A distinguished set of non-normalizable modes can be selected by requiring that they have regular behavior as $y \to 0$ or $z \to i\pi/4$, namely that they are single valued in the complex coordinate plane just like the normalizable ones.

Using the transformation formulae for hypergeometric functions we find, as a function of $y$, 
\begin{multline}
  \Psi_{\rm non-norm}~~ \sim (-1)^{-\omega/2 + h_-}
  {\Gamma(2h_{-}) \Gamma({\omega + im \over 2}) \over \Gamma(h_- +{i m \over 2})
  \Gamma(h_-+{\omega \over 2})} {}_2F_1(-{\omega \over 2} + h_-, 
  -{\omega \over 2} + h_+;  1 - {{\omega+im} \over 2}; y) \\
  + (-1)^{h_- + im/2}{\Gamma(2h_-) \Gamma(-{\omega + im \over 2}) \over 
  \Gamma(-{\omega \over 2}+h_-)\Gamma(h_- - i {m \over 2})}~
  y^{({\omega + im \over 2})}{}_2F_1(h_- + i{m\over 2}, h_++ i{m\over 2};
  1+ i{m\over 2}+{\omega\over 2}; y) ~.
 \label{eq:notz0}
\end{multline}
The second term again implies a cut in the complex $z$ plane starting from $y =0$.   For special values of $\omega$, namely $\omega = - 2 h_- - n ,~~ n =0,1,2,\cdots$ the second term vanishes, but we need to include non-normalizable modes for all $\omega$ as discussed above.
Notice, however, that for general $\omega$ the cuts appearing in (\ref{eq:notz0}) are identical in form to the cuts in the normalizable modes (\ref{eq:z0}).  Therefore, we will be able to cancel the cuts in (\ref{eq:notz0}) by adding a suitable linear combination of normalizable modes. 
Indeed, the following combination of non-normalizable and normalizable modes is regular at the origin:
\begin{equation}
\Psi_{\rm non-norm}^{\rm regular} = \Psi_{\rm non-norm}-(-1)^\nu  {{\Gamma(2h_-) \Gamma(-{\omega \over 2}+h_+) \Gamma(h_+-{im \over 2})} \over {\Gamma(2h_+) \Gamma(-{\omega \over 2}+h_-) \Gamma(h_--{im \over 2})}} \Psi_{norm}
\label{linearcombo}
\end{equation} 
where $\nu =h_+-h_-$ and $\Psi_{\rm non-norm}$ and $\Psi_{\rm norm}$ are as in \eqref{eq:solnotnorm} and \eqref{eq:solnorm}.

\subsubsection{The case of integer $\mathbf{\nu=h_+-h_-=\sqrt{1+\mu^2}}$ and a slightly different approach}
\label{sec:different}

Perhaps a simpler way to obtain the basis of normalizable and non-normalizable solutions discussed here is to start by picking the regular solution to \eqref{eq:hyper} near $y=0$. For a generic $\omega$, this solution will be non-normalizable. However, for certain quantized frequencies, the solution should become normalizable. The general solution to \eqref{eq:hyper} is given by
\begin{equation}
\chi(y)=c_1 \,  F(A,B;C;y)+ c_2 \, y^{1-C} F(A-C+1,B-C+1;2-C;y)
\label{generalsol}
\end{equation}
where $A, ~B$ and $C$ are given in \eqref{case1}. For this solution to be regular at $y=0$, we need the second term to be absent, {\em i.e.} $c_2=0$. Then, starting with the regular solution,  
\begin{equation}
\chi(y)=F(-{\omega \over 2}+h_-, -{\omega \over 2}+h_+; 1 - \Bigl( {\omega +im \over 2 } \Bigr), y)
\label{regularsol}
\end{equation}
we use the transformation formula \eqref{transformation} to determine its behavior near $y=\infty$:
\begin{eqnarray*}
\chi(y)&=&{\Gamma \left(1 - \Bigl( {\omega +im \over 2 } \Bigr)\right) \Gamma(h_+-h_-) \over \Gamma(-{\omega \over 2}+h_+)  \Gamma(h_+  - { im \over 2})} (-y)^{-{\omega \over 2} +h_-} F(-{\omega \over 2}+h_-, h_-+{im \over 2}; 2h_-; {1 \over y})  \\
& & ~~~
 +{\Gamma\left(1 - \Bigl( {\omega +im \over 2 } \Bigr)\right) \Gamma(h_--h_+) \over \Gamma(-{\omega \over 2}+h_-)  \Gamma(h_- - { im \over 2})} (-y)^{-{\omega \over 2} +h_+}F(-{\omega \over 2}+h_+, h_++{im \over 2}; 2h_+; {1 \over y})
\end{eqnarray*}
The first term corresponds to the non-normalizable piece and for $-{ \omega \over 2}+h_+=-n$, where $n$ is a non-negative integer, $\Gamma(-{ \omega \over 2}+h_+)$ in the denominator has a pole, implying that for these particular quantized frequencies, the solution only has a normalizable piece. This normalizable solution as well as the quantization condition on $\omega$ is, of course, the same as what we obtained earlier  in \eqref{eq:solnorm}. Also, for a generic $\omega$, it is easy to see that \eqref{regularsol} is precisely the linear combination \eqref{linearcombo} which is non-normalizable and regular at the origin. 

However, this analysis and our previous discussions break down when $\nu=h_+-h_-$ is an integer.  In this case,   the transformation formula \eqref{transformation} is not valid.  In fact, we then use the more complicated formula 
\begin{eqnarray*}
  &&\chi(y)=F(A,B;C;y)={\Gamma(C) (-y)^{-B} \over \Gamma(B) \Gamma(C-A)} 
  \sum_{n=0}^\infty {(A)_{n+\nu} (1-C+A)_{n+\nu} \over n! (n+\nu)!} y^{-n}\bigg\{ \ln(-y)+\psi(1+\nu+n) \\ 
  & &~~~+ \psi(1+n)-\psi(B+n)-\psi(C-B-n) \bigg\} +(-y)^{-A} { \Gamma(C) \over \Gamma(B)} \sum_{n=0}^{\nu-1} 
  { \Gamma(\nu-n) (A)_n \over n! \Gamma(C-A-n)} y^{-n}
\end{eqnarray*}
where $A,~B$ and $C$ are given in \eqref{case1} and $\psi(z)=d \ln \Gamma(z)/dz$ is the digamma function. The second term with 
leading power $y^{-A}$ corresponds to the non-normalizable piece, which vanishes when $\Gamma(B)$ has a pole, {\em i.e.}  $-{ \omega \over 2}+h_+=-n$ 
is a non-negative integer. For this subset of frequencies, the analysis is more subtle since the first term might naively seem to vanish due to the
$\Gamma(B)$ factor appearing in the denominator. However, notice that $\Psi(B+n)$ is also singular for this same subset of frequencies, and a careful 
limit gives the same result as before \eqref{eq:solnorm} for the normalizable modes.

\subsection{Green functions from the method of images}
\label{sec:images}

The non-normalizable modes obtained in the previous section give the bulk to boundary propagator,  Fourier transformed with respect to $t$ and $\phi$. 
We can also obtain the bulk to boundary propagator from a certain scaling limit of the bulk Feyman propagator on the orbifold.   
We can compute the Feynman Green function for scalar fields on the orbifold  
 by the method of images: {\em i.e.} we start with the Feyman propagator in $\ads{3}$ and then add all images under the orbifold action.   
This method implies a certain choice of vacuum on the orbifold that descends from the unique 
$\SL(2,\bR) \times \SL(2,\bR)$ invariant vacuum on $\ads{3}$.  The Feynman Green function on global $\ads{3}$ \cite{Greenfunc} is
\begin{equation}
  -iG_F(x,x^\prime) = \frac{1}{4\pi R}\left(z^2 -1\right)^{-1/2}\left[z+\left(z^2-1\right)^{1/2}
  \right]^{1- 2h_+} \,,
 \label{eq:2pt}
\end{equation}
where $z=1+R^{-2}\sigma(x,\,x^\prime) + i\epsilon$, $\sigma(x,\,x^\prime)$ being the invariant
distance among two points in the embedding space $\bR^{2,2}$.  That is, $\sigma(x,\,x^\prime)=
\frac{1}{2}\eta_{MN}(x-x^\prime)^M(x-x^\prime)^N$ and for $\mu^2 > 0$
\[
2h_+ = 1 + \sqrt{1 + \mu^2} \, .
\]
Thus, the Green function on the orbifold will be given by 
\begin{equation}
  -iG_{F/\Gamma}(x,x^\prime) = -i\sum_{n} G_F(x,x^\prime_n) =
  \frac{1}{4\pi R}\sum_{n}\left(z_n^2 -1\right)^{-1/2}\left[z_n+\left(z_n^2-1\right)^{1/2}
  \right]^{1-2h_+}
 \label{eq:2ptorb}
\end{equation}
where we defined $z_n (x,x^\prime) = z(x,x^\prime_n)$ and $x^\prime_n$ is the $n$th image point. It 
is easy to work out its value in the adapted coordinate system where the discrete identification is
easy to implement. The answer is 
\[
  z_n = \sin\Delta t \sinh \Delta\phi_n \sinh\tilde\Delta z +
        \cos\Delta t\cosh\Delta\phi_n\cosh \Delta z + i\epsilon\,,
\]
where $\Delta t=t-t^\prime$, $\Delta z=z-z^\prime$, $\tilde\Delta z = z+z^\prime$ and
$\Delta\phi_n = \phi - \phi^\prime + 2\pi n$.

\section{The holographic dual}
\label{holography}

Given an asymptotically AdS spacetime, the usual instruction concerning holography is that each disconnected component of the boundary will contain a 
copy of the canonical field theory dual of AdS space.  These theories may be decoupled, entangled, interacting, identified or otherwise related depending on the 
specific circumstances.  In the eternal BTZ black hole, for example, the two boundaries are thought to contain entangled copies of the D1-D5 CFT
\cite{eternalBH,BKLT}.  By contrast, in $\ads{2}$, which is a strip, there are subtleties concerning the status of the  quantum mechanical dual theories that could live on each boundary. It has been suggested that the theories on these two boundaries should be identified with 
each other \cite{S98,ads2}. In explorations of holography in de Sitter space \cite{dsdual}
similar issues have arisen: if there exists a field theory dual associated to the two de Sitter boundaries, should there be two entangled 
theories or should the two theories be identified?  

In the present case, the spacetime (\ref{eq:fmetric}), seen as an asymptotically $\ads{3}$ spacetime, would appear to be dual to two 
(possibly entangled or interacting) copies of the D1-D5 CFT defined on the two null-cylinder boundaries. As we have seen, the surface at fixed 
radius on the orbifold is a boosted cylinder and the boost approaches infinity as the AdS boundary is reached, creating a null cylinder.  
This suggests the dual contains two DLCQ copies of the D1-D5 CFT. Another approach to duality might be to regard the dual as arising from an 
orbifold of the dual of $\ads{3}$. Interestingly, as we have seen, the isometry under which we identify AdS is a conformal transformation of the boundary.
This suggests that the dual is a ``conformal orbifold" in which a theory is quotiented by a conformal transformation. We will compare and contrast these two approaches to the definition of a dual theory in 
Sec.~\ref{twoapproaches}.

Perhaps the most pedestrian way of approaching holography in AdS is to blindly apply  the well-known prescription for computing correlation functions of a 
dual CFT on the boundary in terms of bulk data \cite{GKPW}.  We have two related ways of approaching this -- we can either use the
bulk-boundary propagator prescription or the on-shell action prescription. Since this is the most concrete 
path towards duality, 
we will take it first in the section below.

\subsection{Two point correlators and the choice of vacuum}

The classic AdS/CFT correspondence \cite{adscft} is defined by equating the partition function for string theory on AdS seen as a functional of 
boundary data to the generating function of correlation functions of the dual field theory \cite{GKPW}.  In the semiclassical limit of the bulk theory 
(large $N$ limit in the dual) this amounts to equating the bulk on-shell action to the boundary generating function of correlators.  
In our case there are two disconnected components to the boundary and the non-normalizable mode solutions whose boundary values are dual sources diverge 
on both boundaries. As we will see, this means that the on-shell action always has contributions from both boundaries.  This suggests that the sources in the components of the dual theory on the two boundaries have correlated sources, but we will see that the situation is more subtle and interesting than the naive expectation, and that it is in fact possible to compute independent correlation functions within each boundary as well as between them.

Another approach to computing correlation functions of the dual to AdS space is to use a ``bulk-boundary propagator'' $G_{B\partial}$ in terms of which a diagrammatic expansion computes CFT correlators.
 We can define $G_{B\partial}$ by taking the boundary limit
of the bulk Feynman propagator.  From this perspective, it is possible to obtain a propagator from a single boundary into the bulk, in terms of which correlation functions of a dual defined on a single boundary or between boundaries can be obtained.   We will compare these two approaches to correlation functions below.

Another tricky issue here is that in general, the basis of non-normalizable modes is ambiguous because one can add normalizable modes to the former without changing the growth near infinity.  In \cite{BKLT,tommy} it was pointed out the choice of added  normalizable modes has a bearing on the choice of vacuum for the AdS/CFT correspondence since the choice of non-normalizable modes in effect specifies the Fourier transform of the bulk-boundary propagator of the correspondence.  It was also explained in \cite{BKLT,tommy} that requiring the bulk-boundary propagator to arise from Euclidean continuation, namely as an analog to the bulk-bulk Feynman propagator, uniquely selected the normalizable component in the non-normalizable mode. 
Our spacetime does not have a Euclidean continuation.  However, a distinguished basis of non-normalizable modes can be obtained by canceling the cut that appears 
in \eqref{eq:notz0} as $y \to 0$ by addition of a normalizable mode \eqref{eq:z0}. Below, we explore this issue which is related to the choice of vacuum for 
the orbifold.

\subsubsection{Correlators from the method of images} 
\label{sec:imagesa}

First we make use of the expression \eqref{eq:2ptorb} for the Feynman Green function on the orbifold $G_{F/\Gamma}$  to derive the bulk-boundary propagator.
We use the notation $G_{B\partial \pm}$ for the propagator from the boundaries at $z \rightarrow \pm \infty$ to the bulk.
To evaluate $G_{B\partial +}$ we send $z\to\infty$ in $G_{F/\Gamma}(x,x')$ \eqref{eq:2ptorb} and rescale
by $(e^{2h_+ z})$ to remove the standard asymptotic AdS falloff:  
\begin{equation}
  -iG_{B\partial + }(x,x^\prime) = \frac{1}{2\pi R} \sum_{n} \left(\cos\Delta t\cosh\Delta\phi_n e^{-z^\prime}
  + \sin\Delta t\sinh\Delta\phi_n e^{z^\prime}\right)^{-2h_+}\,,
 \label{eq:bb1}
\end{equation}
where we omitted the $i\epsilon$ prescription.  Similarly,
\begin{equation}
  -iG_{B \partial -}(x,x^\prime) = \frac{1}{2\pi R} \sum_{n} \left(\cos\Delta t\cosh\Delta\phi_n e^{z^\prime}
  - \sin\Delta t\sinh\Delta\phi_n e^{-z^\prime}\right)^{-2h_+}\,.
 \label{eq:bb2}
\end{equation}
Notice that  both bulk-boundary propagators diverge
as $z^\prime \to \pm \infty$ when $\Delta t $ or $ \Delta \phi_n$ equals 0 (or indeed 
suitable multiples of $\pi$)
but otherwise vanish in this limit.   This confirms that these propagators diverge along the boundary lightcones as they should and vanish otherwise.   (The additional singularities at integer multiples of $\pi$ are explained below in terms of the periodicities of the bulk geodesics.)

We can use $G_{B\partial \pm}$ to compute correlation functions within the boundaries 
at $z \to \pm \infty$ and between them by taking the remaining bulk point to the appropriate boundary 
and 
rescaling again by  $(e^{2h_+ z})$.  We will only look at the correlator for nonzero $\Delta t$ and $\Delta \phi_m$ since we are not interested
in contact terms.   Using $G_{\pm\pm}(\hat{x},\,\hat{x}^\prime)$ to denote the 
possible two-point correlators, with $\hat{x},\,\hat{x}^\prime$ being points in the boundaries,
we find:
\begin{eqnarray}
  -iG_{++}(\hat{x},\,\hat{x}^\prime) & = &\frac{1}{2\pi R}\left(\sin\Delta t\right)^{-2h_+}
  \sum_{n} \left(\sinh\Delta\phi_n\right)^{-2h_+}\,, 
 \label{eq:b1b1} \\
  -iG_{--}(\hat{x},\,\hat{x}^\prime) & = &\frac{1}{2\pi R}(-1)^{-2h_+}
  \left(\sin\Delta t\right)^{-2h_+} \sum_{n} \left(\sinh\Delta\phi_n\right)^{-2h_+}\,, 
 \label{eq:b2b2} \\
  -iG_{+-}(\hat{x},\,\hat{x}^\prime) = -iG_{-+}(\hat{x},\,\hat{x}^\prime) & = &
  \frac{1}{2\pi R}\left(\cos\Delta t\right)^{-2h_+}
  \sum_{n} \left(\cosh\Delta\phi_n\right)^{-2h_+}\,. 
 \label{eq:b1b2}
\end{eqnarray}
We can generate \eqref{eq:b1b2} from \eqref{eq:b1b1} by
shifting $\Delta t\to \Delta t + \frac{\pi}{2}$ and $\Delta \phi_n\to \Delta \phi_n + i\pi/2$ (up to
a constant phase).  This agrees nicely with the fact that we can reach the second boundary from the first via an excursion in the complex plane as described in Sec.~\ref{excursions}.
Notice that the $G_{++}$ and $G_{--}$ are singular when $\Delta t = 0$ or $\Delta \phi_n = 0$.  This
makes sense because $\phi$ and $t$ are lightcone coordinates on the $z \to \pm \infty$ boundaries.
By contrast $G_{+-}$ is singular when $\Delta t = \pi/2$.  This agrees with our analysis of geodesics: there is a null geodesic connecting points on opposite boundaries that are separated by $\Delta t = \pi/2$.   Null geodesics can bounce from one boundary to the other and back again in time $\Delta t =n \pi$,
accounting for the periodicity in the singularities of $G_{++}$. (Strictly speaking, for such integrally spaced intervals $\Delta t$ the limiting procedure leading to (\ref{eq:b1b1}) from (\ref{eq:bb1}) should be revisited 
since the second term with the parenthesis in (\ref{eq:bb1}) vanishes in these case.)

It is also instructive to calculate the correlation function in momentum space.  The Fourier transform of $G_{++}$ is:
\begin{equation}
\tilde{G}_{++}(\omega, m) = 
\int_{-\infty}^\infty dt  \int_0^{2\pi} d\phi \, \,  e^{i \omega t} e^{i m \phi} G_{++}(t,  \phi).
\end{equation}
This is given by (we refer the reader to appendix \ref{sec:fourier} for details):
\begin{multline}
  \tilde{G}_{++}(\omega,m) =  {i \over 2\pi R} \, (-1)^{-2h_+}\frac{1}{4}\,\left(\frac{i}{2}\right)^{2h_+-1} \frac{1}{2^{2h_+}}\,\left(\Gamma (1 - 2h_+)\right)^2\,
  \frac{\Gamma (h_+ - im/2)}{\Gamma (h_- - im/2)}\,\frac{\Gamma (-\omega/2 + h_+)}{\Gamma (-\omega/2 + h_-)} \\
\times \, 
  \left(e^{i\theta} + (-1)^{-2h_+}\right)\left\{ \frac{\sin (\pi(\omega/2 + h_-))}{\sin (\pi(\omega/2+h_+))}+ (-1)^{-2h_+}\right\}~.
 \label{fourierfinal}
\end{multline}
where 
\begin{equation}
 {\csc\Bigl( \pi ( h_++{im \over 2}) \Bigr) \over \csc\Bigl( \pi ( h_- +{im \over 2}) \Bigr)}=e^{i \theta} \, .
\end{equation}
This form of the correlation function will be useful for comparison with the on-shell action.

\subsubsection{Correlation functions from boundary variation of the on-shell action}
\label{sec:bulkaction}

To compute dual CFT correlators from AdS action calculations we must choose a basis of non-normalizable modes.    Since we can always add a normalizable piece to such a solution, the most general basis is 
\begin{equation}
\bar\Psi  (z,\,t,\,\phi) = e^{-i\omega\,t+i\,m\,\phi}\,\bar\Psi (z,\,\omega,\,m) =
    e^{-i\omega\,t+i\,m\,\phi}\left(\psi_{\text{non-norm}}(z,\,\omega,\,m) +
    C(\omega,\,m)\,\psi_{\text{norm}}(z,\,\omega,\,m)\right).
\label{eq:linear}
\end{equation}
Since $\bar\Psi$ can be regarded as the Fourier transform of the bulk-boundary propagator,
the frequency dependent coefficient
 $C(\omega,m)$ is related to the choice of vacuum implied by this basis.
A distinguished basis of non-normalizable modes is determined by canceling the cuts in the solution as $y\to 0$ in \eqref{eq:solnotnorm} by the addition of a normalizable 
component.   This requires that:
\begin{equation}
    C(\omega,\,m) = -(-1)^{\nu} \frac{\Gamma (2h_-)\,\Gamma (-\frac{\omega}{2} + h_+)\,\Gamma (h_+ - i\frac{m}{2})}{
    \Gamma (2h_+)\,\Gamma (-\frac{\omega}{2} + h_-)\,\Gamma (h_- - i\frac{m}{2})}
\equiv \CC(\omega,m)~.
\label{specialC}
\end{equation}

The on-shell action is given by
\begin{equation}
  S_{\rm bulk} = \lim_{\Lambda \to \infty} {1 \over 2}\cosh\,2\Lambda \Bigl(\int dt d\phi\, \bar\Psi \partial_z \bar\Psi |_{z=\Lambda}- 
  \int dt d\phi\, \bar\Psi \partial_z \bar\Psi |_{z=-\Lambda}\Bigr)~,
 \label{eq:bulka}
\end{equation}
after integrating by parts and using the equations of motion in the metric \eqref{eq:fmetric}.  Writing
\begin{equation*}
  \bar\Psi  (z,\,t,\,\phi) = \frac{1}{2\pi}\int d\omega \sum_m\,e^{-i\omega\,t+i\,m\,\phi}\,
  \bar\Psi (z,\,\omega,\,m)~.
\end{equation*}
in momentum space, the on-shell action becomes
\begin{equation}
  S_{\rm bulk} = \lim_{\Lambda \to\infty}{\cosh2\Lambda \over 2}
\int d\omega \sum_m
(  \bar\Psi (z,\omega,m)  \partial_z  
  \bar\Psi(z,-\omega,-m) |^{z=\Lambda}_{z=-\Lambda} ~.
 \label{eq:bulkb}
\end{equation}

It is convenient to rewrite this  in terms of a bulk-boundary propagator ${\cal K}(z,m,\omega)$ which is normalized to
approach $1$ at the boundary.   
Since we have two boundaries here we must choose which one to normalize $\CK$ on.  Choosing 
$\CK \to 1$ as $z \to \Lambda$ (where we take $\Lambda \to \infty$ at the end) 
we can write
\begin{equation}
  {\cal K}_+(z,\,\omega,\,m) = \frac{\bar\Psi (z,\,\omega,\,m)}{\bar\Psi (\Lambda,\,\omega,\,m)}~.
 \label{eq:bbpropa}
\end{equation}
In terms of $\CK_+$ a non-normalizable mode $\bar\Psi (z,\,t,\,\phi)$ whose Fourier components as $z\to\Lambda$ are $\bar{J}_+(\omega,\,m)$ 
can be written as
\begin{equation}
   \bar\Psi (z,\,\omega,\,m) = \bar{J}_+(\omega,\,m)\,{\cal K}_+(\Lambda,\,\omega,\,m)~.
\end{equation}
Using this in the on-shell action \eqref{eq:bulkb} gives
\begin{equation}
  S_{\rm bulk} = \lim_{\Lambda \to\infty}{\cosh 2\Lambda \over 2} \int d\omega \sum_m \bar{J}_+ (\omega,\,m)\,\bar{J}_+(-\omega,\,-m)\,
  {\cal K}_+(z,\,\omega,\,m)\,\partial_z\,{\cal K}_+(z,\,-\omega,\,-m)|^{z=\Lambda}_{z=-\Lambda} ~.
 \label{eq:bulkc}
\end{equation}

The boundary two-point function  evaluated in momentum space should then be
\begin{equation}
G^\prime_{++}(\omega,m)
 = \frac{\delta S_{\rm bulk}}{\delta \bar{J}_+ (\omega,\,m)\,
  \delta \bar{J}_+ (\omega^\prime,\,m^\prime)}~.
\label{twopointdef}
\end{equation}
By construction we are considering the two-point function in the boundary at $z \to \infty$ since we specified the boundary data there; hence the notation $G_{++}^\prime$.  Note however
that the boundary data at $z \to -\infty$ are fully specified also since
\begin{equation}
  {\cal K}_+(-\Lambda,\,\omega,\,m) \to e^{-m\pi/2}\,e^{i\pi(\omega/2 + h_-)} \quad \text{as} \quad \Lambda\to\infty~.
\label{Kpluslimit}
\end{equation}
Thus the two-point function picks up a contribution from both boundary terms (\ref{eq:bulkc}) in the bulk action.

Computing the two-point function from \eqref{eq:bulkc} and \eqref{twopointdef} by explicitly using $\CK_+$ is a tedious but straightforward exercise.    We spare the reader the details,
but recall that the basic philosophy, following \cite{mathuretal} is to evaluate the action in a power series in $e^\Lambda$, drop singular contact terms, and extract the resulting contribution to the 2-point function.   The leading term that contributes always arises from a interaction between the non-normalizable and the normalizable pieces of $\bar\Psi$ and hence of $\CK$, and can readily be identified by the 
 requirement of conformal invariance, which fixes the scaling with the cut-off $\Lambda$ to be $\left(e^\Lambda\right)^{2-2\nu}$, where $\nu=h_+ -h_-$.
Carrying out this computation with $h_- < 0$ and $h_+ > 0$, and taking the functional derivatives, 
we obtain, after some suffering, a boundary 2-point function of the form.
\begin{equation}
G^\prime_{++}(\omega,m)
= \delta (\omega + \omega^\prime)\,\delta_{m+m^\prime} \, \left( 
 A_+ \, C(\omega,\,m) +  A_- \, C(-\omega,\,-m) 
\right)~,
 \label{eq:2ptbulkaction}
\end{equation}
where the coefficients $A_\pm$ could depend on $\nu = h_+ - h_-$, $m$ and $\omega$. Both terms have  contributions from both boundaries.     Using (\ref{asadlove}) we can rewrite this as
\begin{equation}
G_{++}^\prime(\omega,m) = C(\omega,m) \, Q^\prime(\nu,\omega,m)
\label{onshellagain}
\end{equation}
where $Q^\prime$ contains numerical coefficients and various trigonometric functions of  the arguments.
Note that, as described in Sec.~\ref{sec:different}, when $\nu$ is an integer the non-normalizable mode contains logarithms and so some further effort is necessary in the computations.

\subsubsection{Comparison of the two methods}

First of all consider the two-point function computed from the sum-on-images (\ref{fourierfinal}).  Using
\begin{equation}
 \frac{1}{\Gamma (2h_+)} = -\frac{1}{\pi}\Gamma (-\nu)\,\sin (\pi\,\nu)
~~~;~~~
  \Gamma (2h_-) = (-\nu)\,\Gamma(-\nu) ~~~ \Longrightarrow ~~~
\Gamma(-\nu)^2 = {\Gamma(2h_-) \over \Gamma(2h_+)}  {\pi \over \nu  \sin{\pi \nu}} 
\end{equation}
it is easy to show that the sum-on-images result is
\begin{equation}
\tilde{G}_{++}(\omega,m) = \CC(\omega,m) \, Q(\nu,\omega,m) 
= -\frac{\Gamma (2h_-)\,\Gamma (-\frac{\omega}{2} + h_+)\,\Gamma (h_+ - i\frac{m}{2})}{
   \Gamma (2h_+)\,\Gamma -\frac{\omega}{2} + h_-)\,\Gamma (h_- - i\frac{m}{2})}  \, (-1)^{\nu} Q(\nu,\omega,m) 
\end{equation}
where $\CC$ is the special ratio of Gamma functions appearing in (\ref{specialC}) and $Q$ is a 
combination of numerical factors and trigonometric functions.  This gives a result of the same form as the on-shell action computation (\ref{onshellagain}) provided we pick a basis of non-normalizable modes  in (\ref{eq:linear}) with a normalizable component weighted by $C(\omega,m) = \CC(\omega,m)$.  As
stated before, the choice of $C$ amounts to a choice of vacuum, and setting $C=\CC$ picks a vacuum 
in which possible cuts in the complex $z$ plane for bulk-boundary propagator are absent.
 When $2h_+$ is an integer all the formulae are somewhat modified because of cancellations between factors that go to zero and to infinity in this limit, and because of the appearance of log terms in the non-normalizable modes.   We do not investigate the details of this here since experience with the AdS/CFT correspondence has shown that correlation functions do not change qualitatively in the integer limit.

\paragraph{One boundary versus two: }
In the sum-on-images calculation it was clear that by taking different boundary limits of the bulk Feynman propagator we could calculate different boundary correlators, $G_{\pm\pm}$, between points on the same boundary and on opposite boundaries.
The on-shell calculation initially appears to have a rather different character.  As we discussed, the bulk non-normalizable modes diverge on both boundaries at the same time.  A bulk-boundary propagator derived from them and normalized to $1$ as $z \to \infty$ behaves as (\ref{Kpluslimit}) as $z \to -\infty$.  
This suggests that sources for the dual CFT must be turned on in a correlated way in both boundary components.    Transforming to position space, switching on a source $J_+(t,\,\phi)$ on the $z \to \infty$ boundary appears to switch on a source
 $J_-(t,\phi)= e^{i\pi\,h_-}\,J_+(t-\frac{\pi}{2},\,\phi +i \frac{\pi}{2})$
 in the $z \to -\infty$ boundary.   Thus, one of the sources is turned on at a point that is shifted into the complex coordinate plane for the boundary theory.  Sources at {\it real} coordinates appear to be independent of each other.
Nevertheless, the correlation between sources on the two sides suggests that the dual need only consist of a theory on one of these boundaries.    
However, this is a misleading intuition as we learn by comparing with the BTZ black hole.  The authors of \cite{esko,eskoper} compute the propagator from one of the two BTZ boundaries to a point in the bulk in the same asymptotic region and find:
\begin{equation}
K_{{\rm BTZ}}(r,u_+,u_-;u_+^\prime,u_-^\prime) = 
\sum_{n=-\infty}^\infty
{
 \left(
{r_+^2 - r_-^2 \over r^2}
 \right)^{h_+} \, 
e^{-2\pi h_+ [T_+ \, \Delta u_+ + T_- \, \Delta u_- + (T_+ + T_-) 2\pi n]}
\over
\left\{
{r_+^2 - r_-^2 \over r^2} + ( 1 - e^{-2\pi T_+ [ \Delta u_+ 2\pi n]})
( 1 - e^{-2\pi T_- [ \Delta u_- 2\pi n]})
\right\}^{2h_+}
}
\label{BTZbulkbound}
\end{equation}
Here $r$ is the radial coordinate of the BTZ black hole, $r_\pm$ are the coordinates of the outer and inner horizons, $u_\pm$ and $u_\pm^\prime$ are lightcone coordinates in the
bulk and boundary respectively, and $T_\pm$ are temperature parameters.
As explained in \cite{eskoper} points can be transported from one asymptotic region of the eternal geometry to another by the transformations
\begin{equation}
T_\pm u_\pm \to T_\pm u_\pm \mp {i \over 2}
\label{BTZtransform}
\end{equation}
which are reminiscent of the discrete complex transformations in Sec.~\ref{excursions}.  Note that if $\Delta u_\pm = 0$, $K_{{\rm BTZ}}$ diverges as $r \to \infty$ as befits 
the Fourier transform of a  non-normalizable mode in the BTZ spacetime.   Applying the transformation (\ref{BTZtransform}) to the bulk point to move it to the other asymptotic region we find that the transformed propagator will diverge as $r \to \infty$ (thus approaching the second boundary) provided $\Delta u_\pm = \pm  i/2$.   
Fourier transforming will thus lead to a non-normalizable mode that diverges at both boundaries.  Alternatively if it is normalized to unity on one boundary it will pick up a momentum dependent phase at the other one.   This is precisely the situation we have in our orbifold.  Thus, just as in BTZ it is expected that we have independent sources and CFTs defined (at least at real coordinate points) on each conformal boundary.

We can follow this procedure from the BTZ black hole as follows.   First,  we have seen that the on-shell action, even expressed just in terms of $J_+$, collects contributions from both boundaries.  If we expressed the action in terms of $J_-$ we would get essentially the same result, agreeing with the fact that $G_{++} \sim G_{--}$ in the sum-on-images calculation.
To compute $G_{+-}$ from the bulk action, a natural prescription is to express one of the fields $\bar\Psi$ in the on-shell action (\ref{eq:bulka},\ref{eq:bulkb}) in terms of $J_+$ and the other in
terms of $J_-$.
In view of (\ref{Kpluslimit}), it is easy to show that the resulting bulk action will lead to a two point function
\begin{equation}
G^\prime_{+-}(\omega,m) \sim e^{-m\pi/2} e^{i\pi(\omega/2 + h_-)} \, G_{++}^\prime(\omega,m)
\end{equation}
After Fourier transforming to position space we get
\begin{equation}
G^\prime_{+-}(\Delta t, \Delta \phi) \sim G_{++}^\prime(\Delta t - \pi/2, \Delta \phi + i\pi/2)
\end{equation}
This exactly reproduces the shift relationship between the sum-on-images 2-point function $G_{++}$ at $z \to \infty$  and $G_{+-}$ between $z \to \pm\infty$. Thus both the on-shell action
and sum-on-images approaches 
are nominally capable of computing the correlators on either boundary or between them.  The real question is whether these different pieces of data are independent or redundant information in a dual formulation.  We will explore this further below.

\subsection{Orbifold Holography}
\label{twoapproaches}

Above  we approached holography on our orbifold by attempting to define on-shell action and bulk-boundary propagator approaches to computing CFT correlation functions.  Here we explore general properties that the dual must have if (a) it is defined on the null boundary of our orbifold, or (b) it descends from the dual to global $\ads{3}$ by our orbifold action.

\subsubsection{Holography and DLCQ CFTs}

A standard technique for computation in the AdS/CFT correspondence is to truncate 
the space at some fixed large radial coordinate which serves as a cutoff on 
the bulk.  The field theory defined on the boundary surface is a regulated 
version of the CFT dual to the entire space, and sending the boundary to 
infinity corresponds to removing the cutoff.  Let us imitate this procedure 
here.

The metric on fixed $z$ surfaces is $g = l^2(-dt^2 +d\phi^2 + 2 \sinh(2z) dt \, d\phi)$.  As explained earlier,
 this is a boosted version of the usual timelike cylinder with the boost parameter approaching infinity as $|z| \to \infty$.   Thus we expect that the dual theories theories living on each boundary are discrete light cone quantized (DLCQ) since they will have a compact null direction.  The simplest way to work out the spectrum is to consider how the $U(1) \times SL(2,R)$ symmetries are realized on surfaces at fixed $z$.   
We can then work out the spectrum in the limits $z \to \pm \infty$ using representation theory.

As $z \to \infty$, the $SL(2,R)$ symmetry generators (\ref{sl2gens}) and the $U(1)$ associated with $\phi$
become
\begin{eqnarray}
\bar{\CL}_0 &=& {i \over 2} {\partial \over \partial \phi} \nonumber \\
 \CL_0 &=& {i \over 2} {\partial \over \partial t} \nonumber  \\
 \CL_+ &=&  \frac{e^{2it}}{2}\left[\tanh(2z) 
  {\partial \over \partial t} + {1\over \cosh(2 z)} 
  {\partial \over \partial \phi} + i 
  2h_+\right]  \label{boundsl2gens1} \\
  \CL_- &=& -\frac{e^{-i2t}}{2}\left[ \tanh(2z) 
  {\partial \over \partial t} + {1\over\cosh(2 z)} 
  {\partial \over \partial \phi} - i 
 2h_+  \right] \, , \nonumber 
\end{eqnarray}
The constant $2h_+$ appear when the generators acting on a representation with highest weight $h_+$.  From the point of view of the bulk generators (\ref{sl2gens}) this constant appears 
from the action of $\partial_z$ on the asymptotic scaling of the normalizable modes.   The latter three 
generators realize an approximate $SL(2,R)$ symmetry, up to terms of order $e^{-4z}$ which 
are irrelevant in the large $z$ limit. The fact that the $SL(2,R)$ only acts approximately at a fixed $z$ is of course traced to the fact that cutting off the bulk imposes a cutoff in the dual CFT, breaking conformal invariance.
 If $\phi$ had not not been periodic as it is in our orbifold, there would have been a second approximate $SL(2,R)$ symmetry on surfaces of fixed $z$, generated by the first of the 
four operators above, and by two more obtained by exchanging $t$ with $i\phi$ and $\phi$ with $-it$ in $\CL_+$ and $\CL_-$.    However, the resulting exponential dependence on
$\phi$ leads to generators that are not well-defined on the orbifold since $\phi \sim \phi + 2\pi$.  Thus the second possible $SL(2,R)$ does not  exist even in an approximate sense at any fixed $z$.
So as $z \to \infty$ the symmetry generators of the dual as inherited from the bulk are:
\begin{eqnarray}
\bar{\CL}_0 &=& {i \over 2} {\partial \over \partial \phi} \nonumber \\
 \CL_0 &=& {i \over 2} {\partial \over \partial t} \nonumber  \\
 \CL_+ &=&  \frac{e^{2it}}{2}\left[
  {\partial \over \partial t} +   i 
2 h_+ \right]  \label{boundsl2gens2}  \\
\CL_- &=&  - \frac{e^{-2it}}{2}\left[
  {\partial \over \partial t} -   i 
2h_+ \right]  \nonumber
\end{eqnarray}
The latter three operators exactly generate the $SL(2,R)$ algebra $[\CL_0,\CL_\pm] = \mp \CL_\pm$ and $[\CL_+,\CL_-] = 2\CL_0$. By simply examining the metric $ds^2 \propto dt \, d\phi$
on the $z \to \infty$ surface  it my seem surprising that two $SL(2,R)$s are not present as boundary reparameterizations of $t$ and $\phi$, but this is because 
of the limiting procedure described above.   A similar analysis applies as $z \to -\infty$.

The spectrum of the dual theory at $z \to \infty$ will carry charges $(\omega,m)$ under $\CL_0$ and $\bar{\CL}_0$.  Since $\phi$ is periodic, $m$ will be an integer and
the corresponding wavefunction will have a factor $e^{i m \phi}$.
A highest representation of the $SL(2,R)$ is given by $\CL_- |0\rangle = 0$, $|n\rangle = \CL_+^n |0
\rangle$, such states will be eigenfunctions of $\CL_0$ with eigenvalue $\omega/2 = h_+ + n$ with $n$ a non-negative integer.  The corresponding wavefunction will have a factor
$e^{-i (2h_+ + 2n) t}$.   The same analysis applies at $z \to -\infty$.

Actually as $z \to \infty$ only negative $m$ are allowed for reasons of unitarity.   This is expected from the DLCQ perspective, 
  but the easiest way to see it here is to observe that both $\phi$ and $t$ are null directions.   Since the metric is $ds^2 \propto dt \, d\phi$, if  we consider 
states with positive $P_t \sim {\rm  Energy}$ 
then for non-tachyonic excitations (i.e. positive effective mass squared) it must follow that
$P_t \, P_\phi \leq 0$.   Since $P_t = 2 \CL_0 > 0$, it follows that $P_\phi = 2\bar{\CL}_0 \leq 0$.  With our conventions that 
the wavefunction is proportional to $e^{i m \phi}$, this implies that $m \geq 0$ for the theory on the $z \to \infty$ boundary.   On the $z \to -\infty$ boundary, the metric becomes $ds^2 \propto - dt \, d\phi$ and so we will similarly conclude that $m \leq 0$.   

Overall the spectrum exactly matches the results from the bulk -- states carry integer $U(1)$ charges and realize a complete $SL(2,R)$.  The novelty, that can be traced to the infinite boost at each boundary, is that the positive (negative) $m$ eigenvalues are realized on different boundaries.  This nicely meshes with the localization of positive (negative) $m$ normalizable modes in the $z>0$ ($z<0$) regions of the bulk.

\subsubsection{Conformal orbifolds of CFTs}
\label{sec:conforb}

Since our spacetime is an orbifold of global $\ads{3}$ we can try to orbifold its CFT dual to construct the field theory describing our spacetime.  
Recall again that $\ads{3}$ has an $SL(2,R)_L \times SL(2,R)_R$ isometry group that is the same as the conformal group of the dual 2d CFT.  
We chose a basis of generators $\xi^{\pm}_i$ for this isometry group in  eqs. (\ref{bulksl2gens1}) and (\ref{bulksl2gens2}).   
Explicit expressions for these generators in terms of the global $\ads{3}$ coordinates (\ref{eq:gmetric})   are easily obtained by
comparing our isometry generators with those defined in \cite{BKL} (see Sec.~4, especially Sec.~4.3 of that paper).  We find that
\begin{eqnarray}
  \xi_1^+ \equiv  -L_1 &=& -\partial_w \\
  \xi_2^+ \equiv L_3 = &=&  \left( {\cosh 2\rho \over \sinh 2\rho } \right) \cos w \, \partial_w 
  - {\cos w \over \sinh 2\rho} \partial_{\bar{w}}
  +  {\sin w \over 2} \partial_\rho\\
  \xi_3^+ \equiv -L_2  &=& \left( {\cosh 2\rho \over \sinh 2\rho } \right) \sin w \, \partial_w 
  - {\sin w \over \sinh 2\rho} \partial_{\bar{w}} -  {\cos w \over 2} \partial_\rho
\end{eqnarray}
where $w = \tau + \theta$ and $\bar{w} = \tau - \theta$ in terms of global coordinates (\ref{eq:gmetric}).
These generators satisfy the $\SL(2,\bR)$ commutation relations $[L_1, L_2] = -L_3$, $[L_1,L_3] = L_2, [L_2,L_3] = L_1$.
The expressions for $\xi^-$ are obtained similarly. Notice that as we approach the $\ads{3}$ boundary $\rho \to \infty$ the generators 
on a fixed $\rho$ surface become
\begin{equation}
  \xi_1^+ \rightarrow -\partial_w ~~~;~~~ \xi_2^+ \rightarrow \cos w \, \partial_w ~~~;~~~ \xi_3^+ \rightarrow -\sin w \, \partial_w
\end{equation} 
which are the standard left-moving $SL(2,R)$ generators of the cylinder.   
(The $\xi^-$ give rise to the right-moving $\SL(2,\bR)$.) The orbifold studied in this paper is an identification of $\ads{3}$ 
by the action of $\xi_2^+$, which is a conformal transformation $\cos w \, \partial_w$ of the boundary.

Orbifolds by a conformal transformation are not very familiar, so it is helpful to study an example: the free boson on the cylinder.    In order to orbifold by the conformal transformation $\cos w \, \partial_w$ we should study states satisfying:
\begin{equation}
e^{i 2\pi \cos w\,\partial_w} \, | s \rangle = | s \rangle
\end{equation}
or equivalently
\begin{equation}
\cos w \, \partial_w \, |s \rangle = k \, |s\rangle ~~~~;~~~~ k \in Z
\end{equation}
The wavefunction for the free boson can be split  into left and right moving pieces $|s \rangle = f(\bar{w}) + g(w)$.   While $f(\bar{w})$ can be any suitable right moving wavefunction,
\begin{equation}
{\partial g \over \partial w} = {k g \over \cos w}
~~~ \Longrightarrow ~~~ g(w) = A \left[ {1 + \sin w \over \cos w} \right]^k
\end{equation}
For $k=0$ the left-moving wavefunction $g(w)$ is constant.    When $k > 0$ ($k < 0$), $g(w)$ is singular at $w = \pi/2$ ($w = - \pi/2$).\footnote{For $k > 0$ $g(w)$ is singular
when the denominator vanishes at $w = \pi/2$.  Both the numerator and the denominator vanish at $w = -\pi/2$, but in this case L'H\^{o}pital's rule shows that $g(w) \rightarrow  0$ as $w \rightarrow -\pi/2$.   Similarly the singularity is at $w = -\pi/2$ for $k<0$.}    
The integer parameter $k$ appearing here should be identified with the integer $U(1)$ charge carried by solutions to the wave equation on the orbifold spacetime since in both cases we are considering eigenstates of $\xi_2^+ \equiv L_3$.  In the DLCQ picture for the dual to the orbifold, the states of positive and negative $k$ appear to be represented separately on the left and right boundary DLCQ theories each of which represents one tower of positive left or right moving momenta.  These sets of momenta, taken together, should be associated with quantum number $k$ appearing above in the conformal orbifold picture of the dual CFT.   
The $SL(2,R)$ arising from the right-movers of the conformal orbifold picture is realized in the DLCQ picture by the $SL(2,R)$ that will act separately on the tower of right or left moving momentum states that survive the DLCQ limit in each boundary.

\paragraph{One CFT or two: Entangled states and interactions}
The interesting picture that emerges from the discussion above is that the dual to our orbifold spacetime can be equivalently thought of as a ``conformal orbifold" of a CFT on a cylinder, or a sum of two DLCQ theories.   The latter theories are separated by the bulk spacetime and the positive and negative $U(1)$ charges of the conformal orbifold appear separately as right/left moving momenta surviving the two DLCQ limits.   (There is some subtlety concerning the states with zero $U(1)$ charge.)   All of this strongly suggests that in the DLCQ picture both boundaries are necessary to construct a complete dual, as in the eternal  BTZ black hole, and unlike some suggestions made in $\ads{2}$ and in de Sitter space.   

A key question is what is the vacuum state for this proposed dual?  In the case of BTZ black holes choosing the Hartle-Hawking vacuum for the bulk spacetime leads to a particular entangled state in the two dual CFTs \cite{eternalBH,BKLT}.    We cannot follow a parallel logic here because of the absence of a Euclidean continuation.  Nevertheless, as we will see in Sec.~\ref{penroselike}, our orbifold can be obtained as a  Penrose-like focussing limit of the BTZ black hole.  The corresponding limit 
of  the BTZ dual will lead to an entangled state in two components of the dual 
to our spacetime.   The discrete excursions in the complex coordinate plane that transport points in one boundary to the other and back again (Sec.~\ref{excursions}, and the resulting complex transformations relating correlation functions on a single boundary (\ref{eq:b1b1})  and between boundaries (\ref{eq:b1b2}), lead to a similar conclusion ).   Indeed, one can easily  write down an entangled state between the two boundary theories that reproduces the required symmetries and relations between 2-point correlators (\ref{eq:b1b1}) and (\ref{eq:b1b2}). 
However, compared to BTZ, the additional challenge here is that the two boundaries are causally connected through the bulk and thus we expect interactions between the two dual components.  In particular, because there are no global horizons in our orbifold, the Hilbert space does not naturally separate into a product with decoupled Hamiltonians acting on each part, although the localization of $m>0$ ($m<0$) modes at $z>0$ ($z<0$) is suggestive.  We are investigating how the appropriate vacuum and interactions can be understood from the dual perspective.

\section{String duality and holography}
\label{stringdualholog}

In Sec.~2 we have seen that the spacetime \eqref{eq:fmetric} arises from at least two different perspectives: (i) as non-singular, causally 
regular discrete quotients of global $\ads{3}$ whose generator is a very particular combination of boosts, and (ii) as an $S^1$ fibration over $\ads{2}$. 
In this section, we will discuss a third inequivalent way of getting our orbifold spacetime: as a Penrose-like limit focusing on the vicinity of the 
horizon of extremal  BTZ black holes.  We will also embed these constructions in string theory, and by using U-duality and liftings to M-theory, we will 
generate several dual descriptions to our original spacetime in type IIA, IIB and M-theory.  Matrix models emerge as the holographic dual 
in several of these dual perspectives, perhaps providing some link with the well-known appearance of matrix models as a dual description of 2d gravity
\cite{2dgrav}.

\subsection{Penrose-like limits of the D1-D5 string and holography}
\label{penroselike}

In type IIB string theory on $T^4$, the near horizon limit of $Q_1$ D1 branes sharing a non-compact direction with $Q_5$ D5 branes wrapped on the  
$T^4$ is $\ads{3}\times S^3\times T^4$. If we compactify the common spatial direction shared by the two sets of branes on a circle of radius $R$, 
and put $n$ units of left or right moving momentum on this circle, the spacetime becomes an extremal 5d black hole whose near-horizon limit is the 
extremal (mass = angular momentum), rotating BTZ black hole times $S^3 \times T^4$. As shown by Strominger and collaborators \cite{lowe,S98,ads2}, the discrete quotient we have considered 
in this paper arises universally as a Penrose-like ``very-near-horizon'' limit of this black hole.    

In terms of the 5d black hole charges  $(Q_1, Q_5, n)$ the near-horizon BTZ metric is
\begin{equation}
  g = l^2T^2\left(dx + \frac{dt}{R}\right)^2 + \frac{U^2}{l^2}\left((R\,dx)^2-dt^2\right)
  + l^2\frac{dU^2}{U^2} \,.
 \label{eq:5bh}
\end{equation}
where ${n\over R}$ is the momentum along the compact direction of the D1-D5 system, $l$ is the AdS scale,  
$T^2=n/(Q_1\cdot Q_5)$, and $x$ is identified as $x\sim x + 2\pi$. 
It is useful to introduce a new radial coordinate
\begin{equation}
  r^2 = \frac{U^2\cdot R^2}{l^2} + l^2\cdot T^2~, \quad r\in[l\,T,\,\infty)~.
 \label{eq:radial}
\end{equation} 
The metric \eqref{eq:5bh} in the new coordinate \eqref{eq:radial} acquires the standard BTZ form, after a rescaling of the timelike
coordinate, $t\to R\,\tau/l$:
\begin{equation}
  g = - \frac{(r^2 - l^2T^2)^2}{r^2\,l^2}\,d\tau^2 + \frac{r^2\,l^2}{(r^2 - l^2T^2)^2}\,dr^2 + r^2
  \left(dx + \frac{l\,T^2}{r^2}\,d\tau\right)^2 ~.
 \label{eq:f5bh}
\end{equation}
In these coordinates, it is clear that the black hole has a horizon at $r_+=l\cdot T$.

The ``very-near-horizon'' limit introduced in \cite{S98}, and also used in \cite{ads2}, 
\begin{equation}
  \frac{U^2\cdot R^2}{l^4\cdot T^2}\to 0~,
\label{verynearhorizon}
\end{equation}
is a focusing limit in the geometry close to the horizon of the extremal black hole, $r\to r_+$, as is easily checked by taking this 
limit in \eqref{eq:radial}.    We can think about this limit in various ways:
$U \to 0$ with $R$ and $T$, $R \to 0$ with $U$ and $T$  fixed, $n \to \infty$
where $n$ is the momentum along the D1-D5 string with all other quantities fixed, etc.  In whichever way we choose to 
think, this is a Penrose-like limit in the sense that we are focusing on a null surface.  Of course, an actual Penrose limit focuses 
on a null {\it geodesic}.  However, we will see some further analogies between these two kinds of limits.

The BTZ geometry described by \eqref{eq:f5bh} is locally $\ads{3}$, and is obtained by discrete identifications of $\ads{3}$.
In the Poincar\'{e} patch description of $\ads{3}$ 
\begin{equation*}
  ds^2 = \frac{l^2}{y^2}\left(d\omega^+\,d\omega^- + dy^2\right)\, ,
\end{equation*}
the identification $x\sim x + 2\pi$ in \eqref{eq:5bh} corresponds to \cite{S98}
\begin{equation}
  \begin{aligned}[m]
    \omega^+ & \sim e^{4\pi T}\,\omega^+ ~,\\
    \omega^- & \sim \omega^- + 2\pi R ~,\\
    y & \sim e^{2\pi T}\,y \,.
  \end{aligned}
 \label{eq:orbpoincare}
\end{equation}
This identification is generated by the following Killing vector:
\[
  \xi = 2\pi\left(R\,\partial_{\omega^-} + T\left(2\omega^+\partial_{\omega^+} +
  y\partial_y\right)\right) \,.
\]
In terms of the generators (see \eqref{bulksl2gens1}) of $\SO(2,2)$ acting linearly in $\bR^{2,2}$ in which $\ads{3}$ is embedded, this Killing vector is 
\begin{equation}
  \xi = 2\pi R\left(J_{02}-J_{01} + J_{13}-J_{23}\right) +
  2\pi T\left(J_{12}-J_{03}\right)\,.
 \label{eq:gorbifold}
\end{equation}
As explained in \cite{S98}, in the ``very-near-horizon'' Penrose-like limit, the geometry is obtained by
keeping only the latter term in the generator (\ref{eq:gorbifold}) of the identification.  
Thus it corresponds to $R \to 0$ or $T \to \infty$ (which implies  $n \to \infty$ since we are are keeping 
$Q_1$ and $Q_5$ fixed in order to fix the AdS scale and the torus moduli) combined with rescalings to normalize the generator. 
In this limit,  we recover the self-dual identification that we have been discussing in this paper since the dominant second term 
in the generator (\ref{eq:gorbifold}) is related to \eqref{eq:killvect} by a rotation in the $\{x^2\,,x^3\}$ plane by $\pi/2$, which is an isometry.

This Penrose-like limit strongly suggests that the dual to our orbifold involves the DLCQ of the D1-D5 string. The momentum along this 
string is $n/R$. Therefore, sending $R \to 0$ or $n \to \infty$ as required by the very-near-horizon limit is equivalent to 
studying the physics in the infinite momentum frame of the D1-D5 string, leading to a DLCQ-like description \footnote{The infinite momentum frame also arose in the context of AdS/CFT for waves propagating in the worldvolume of M2, D3 and M5-branes, by taking the Maldacena decoupling limit keeping the momentum density of the wave fixed \cite{cvetic} .}  
Interestingly, the $R \to 0$ limit achieves this even if the number of quanta of momenta, namely $n$, is fixed.
Since the fixed $n$ geometries realize all the extremal 5d black holes in this compactification, and equivalently 
all the extremal BTZ geometries \cite{stromnearhorizon}, we are studying the properties of the vicinity of the horizon of these black holes.   
We learn the geometry in this region is universal, as is its dual description as a DLCQ theory. It is interesting that the 
extremal 5d black holes appearing here are precisely the ones whose states have been counted in string theory \cite{stromvaf}.  
There is even a limit in which the 4d extremal black holes of string theory whose states can be counted \cite{4dstring} display 
a BTZ in the near-horizon \cite{vijayfinn}.   Thus we are in a sense studying the universal properties of the horizon of all 
black holes whose entropy is understood fully in string theory.

An interesting analogy between this emergence of our orbifold from a focus on a null surface, and the emergence of a pp-wave from 
the focus on a null geodesic \cite{penrose,jose}, is that both spacetimes
have a null boundary \cite{simon}.  In our case the boundary is a null cylinder while the pp-wave 
has a null line boundary.   In the next section we will see that from one perspective the dual to our orbifold is related to a quantum 
mechanical theory, just as the pp-wave should be. A productive methodology in the pp-wave case was to implement the Penrose limit as 
an operation in the dual to the full AdS spacetime, thereby isolating
the sector of the full CFT  that is needed to describe the pp-wave \cite{BMN}.  A useful 
approach here might be to consider an analogous operation focusing on a sector of the dual to the BTZ black hole which has also been  extensively studied.

\paragraph{On the origin of symmetries: }
The first thing to understand is how the $U(1) \times U(1)$ isometry group of the BTZ black hole (generated by the vector fields $\partial_t$ and
$\partial_x$)
 goes over to the $U(1) \times SL(2,R)$ isometry group of the Penrose-like limit, namely our orbifold.
In terms of the Poincar\'{e} generators
\begin{equation}
  \begin{aligned}[m]
    \partial_t & = \frac{T}{R}\left(2\omega^+\partial_{\omega^+} + y\partial_y\right) - \partial_{\omega^-}
\\
    \partial_x & = 2\pi\left(R\partial_{\omega^-} + T\left(2\omega^+\partial_{\omega^+} + y\partial_y\right)\right) \, .
  \end{aligned}
\end{equation}
After the rescaling  $t\to \frac{R}{l}\tau$ the natural ``Hamiltonian'' in BTZ
coordinates is 
\begin{equation}
  \partial_\tau =  \alpha\left(\frac{T}{l}\left(2\omega^+\partial_{\omega^+} + y\partial_y\right) - \frac{R}{l}\partial_{\omega^-}\right)~.
\end{equation}
Thus, in the Penrose-like limit in which $R\to 0$ or $T\to\infty$, $\partial_\tau$ and $\partial_x$ coincide
and give identical quantum numbers.   This $U(1)$ which survives the limit is the compact $U(1)$ generator of our orbifold.
The $SL(2,R)$ that emerges is a new enhanced symmetry appearing in the limit.    This is reminiscent of the enhanced symmetries that can appear in the Penrose limits of AdS space \cite{jose,enhanced}.

\paragraph{On entangled states: }
Eternal non-extremal BTZ black holes have a dual description in terms of two copies of the D1-D5 CFT in an entangled state \cite{eternalBH,BKLT,eskoper}.   Following the conventions of \cite{eskoper}
the temperature $T$ and angular momentum potential $\Omega$ of the non-extremal BTZ can be combined into 
\[
{1 \over T_{\pm}}\equiv {1 \over T} \pm {\Omega \over T}
\]
where $T_{\pm}$ are related to $r_{\pm}$ by
\[
r_{\pm} = \pi l (T_{-} \pm T_{+} )
\]
The non-extremal BTZ is then dual to two decoupled conformal field theories living in an entangled state \begin{equation}
  |\Psi> = \sum_{E,J} e^{-\beta_H (E-\Omega\,J)} |E,J>_1 |E,J>_2 ~,
 \label{eq:entangled}
\end{equation}
where, effectively, we can described the Penrose-like limit by $\Omega\to 1$ and $\beta_H\to\infty$ in the entangled state.\footnote{The maximally extended non-extremal BTZ 
actually has many asymptotic regions if we include the region beyond the singularity.  See \cite{eskoper} for a very interesting discussion of entanglement between dual theories living on all of the asymptotic components.}     The precise state that results then depends
on how $E$ approaches $J$ as $\beta$ diverges.   As we have already described a DLCQ limit is also involved since an infinite momentum limit is effectively being taken at the same time.
  We will not explore this structure any further here, but the appearance of two entangled theories from the Penrose-like  limit justifies again our assertion that that our orbifold is related to an entangled state in two theories.

\subsection{Compactification, two dimensional gravity and quantum mechanics}
\label{compactholg}

As mentioned in section \ref{sec:bulk}, our spacetime \eqref{eq:fmetric} contains a circle of constant radius. The direction
along the circle is a direction of isometry. Thus we can compactify on this circle getting a two-dimensional effective description.
The latter is more easily obtained from the explicit $S^1$ fibration over $\ads{2}$ given in \eqref{eq:fibration}, giving rise to
\begin{equation}
  \begin{aligned}[m]
    g_2 &= -\cosh^2 2z \, dt^2 + dz^2~, \\
    A_1 &=  \sinh 2z \,dt 
  \end{aligned}
\end{equation}
where $A_1$ stands for an electric arising from the off-diagonal component of the 3d metric \eqref{eq:fmetric} and 
$g_2$ is exactly the $\ads{2}$ metric. Thus our spacetime is equally well described as $\ads{2}$ deformed by the addition of a constant
 electric field.  (See \cite{S98} for an earlier discussion of these points.)   Scalar field modes carrying momentum in the $\phi$ direction of our orbifold will carry electric charge on $\ads{2}$ after compactification.  
Because of the background electric field, positive (negative) charges will be attracted towards the $z \to \infty$ ($z \to -\infty$) boundary.  This nicely matches the 
localization of bulk wavefunctions with angular momentum on the orbifold, and also with the splitting of $m > 0$ and $m< 0$ states into the two boundary CFTs.  The boundary of $\ads{2}$ consists of two real lines and 
thus the dual, from this perspective, should be a $0+1$d quantum mechanical theory.  
There has been discussion in the literature concerning whether or not the dual lives on both boundary components.   
Above we saw evidence that both boundaries are needed to describe our spacetime.  Anyway, the dual in our case cannot  be 
simply the dual to $\ads{2}$ because the electric field implies a deformation of the latter. Following the discussion 
in the previous section about the emergence of our spacetime from the D1-D5 string we can conclude that the dual quantum mechanics 
emerges after rescaling a very low energy limit of the dynamics of this string in a sector of fixed but very large angular momentum.    
The appearance of 2d gravity could be a signal of instabilities in our model since back reaction effects are usually very large in 2d, 
leading for example to fragmentation effects which necessitate a sum over geometries \cite{ads2}.  However, if we regard our space as the effective 
description of the vicinity of the black hole horizon in an asymptotically flat spacetime or of the BTZ horizon, then this issue can be disregarded.  
The connection with 2d gravity suggests the possibility that a Matrix model is somehow involved and the appearance of a DLCQ from the 3d perspective 
gives similar indications.

\subsection{Dual backgrounds}
\label{sec:duals}

The orbifold of $\ads{3}$ we are considering has a compact direction. Thus when embedded in string theory, it is natural to study T-duality 
transformations along it and/or the lift to M-theory of this background. 

We will start with the $\ads{3}\times\sph{3}\times\bT^4$
that is obtained as the near horizon limit of a system of D1 and D5-branes. We will follow the
conventions defined in \cite{boris}. Thus all coordinates are dimensionless, the length units begin carried by the components
of the metric. The  fluxes are also dimensionless, which
means that these fields have been normalized in such a way that there are different powers
of $\alpha^\prime$ in the effective supergravity action.

We start with the near horizon limit of the D1-D5 system \cite{MS}
\begin{equation}
  \begin{aligned}[m]
    g & = l^2\left(g_{\ads{3}} + g_{\sph{3}}\right) + \alpha^\prime\sqrt{\frac{Q_1}{v\,Q_5}} dx^i\,dx_i ~,\\
    F_3 &= d\,C_2 = Q_5 \left(\dvol\,\sph{3} + \star_6\dvol\,\ads{3}\right)~, \\
    e^{-2\Phi} &= \frac{v\,Q_5}{g_s^2\,Q_1}\,,
  \end{aligned}
 \label{eq:rr}
\end{equation}
where $l^2 = g_s\,\alpha^\prime \sqrt{Q_1\cdot Q_5/v}$. $Q_1$ and $Q_5$ stand for the charges of the original D1 and D5-branes, 
whereas $v$ determines the volume of the 4-torus. In particular, we are considering a square torus in which $x^i\sim x^i 
+ 2\pi (\alpha^\prime)^{1/2} v^{1/4}$ $i=1,2,3,4$. To describe the self-dual orbifold, we just  replace $g_{\ads{3}}$ 
by the metric appearing in \eqref{eq:fmetric} setting $l=1$. The RR flux, in adapted coordinates \eqref{eq:adapted} is 
\begin{equation}
  F_3 = Q_5\left(\dvol\,\sph{3} + \beta\cosh 2z\,dt\wedge d\phi\wedge dz\right)~.
\end{equation}
In particular, its potential $C_2$ has non-trivial components $C_{t\phi} = \beta\,Q_5\sinh 2z$\footnote{The orbifold discussed
in Sec.~\ref{sec:setup} corresponds to $\beta=1$. It is straightforward to introduce a free parameter, corresponding to the rapidity of the boosts defining the orbifold action by rescaling the coordinate $\phi$.}.

For this classical solution to be reliable,  the string coupling constant should be small. Also, for small $\alpha'$ corrections, we require 
that the radius be bigger than the string scale, otherwise we should use the T-dual description. These two conditions
are summarized below
\begin{eqnarray}
  \frac{v\,Q_5}{g_s^2\,Q_1} & > & 1 \quad \text{(weak coupling)} \label{weakcoupling} \\
  g_s\left(\frac{Q_1\,Q_5}{v}\right)^{1/2} & > & 1 \quad (R_{\text{eff}}^2 > \alpha^\prime)~ \label{alpha}.
\end{eqnarray}
Let us explore the different descriptions of our spacetime as we vary the coupling and the effective radius of the compact dimension.

\paragraph{S-duality:}
At strong coupling, the S-dual configuration is more reliable. 
After the S-duality transformation, the string coupling constant will just be inverted $(\Phi\to -\Phi)$, the RR two-form is interchanged
with the NS-NS two-form potential $(B_2)$ (with a minus sign) and the metric in the string frame is just obtained
from requiring the metric on the Einstein frame to be invariant under the transformation. The final S-dual
configuration is summarized below :
\begin{equation}
  \begin{aligned}[m]
    g & = \alpha^\prime\,Q_5 \left(g_{\ads{3}} + g_{\sph{3}}\right) + \frac{\alpha^\prime}{g_s} dx^i\,dx_i ~,\\
    H_3 &= - \alpha^\prime\,Q_5 \left(\dvol\,\sph{3} + \star_6\dvol\,\ads{3}\right)~, \\
    e^{-2\Phi} &= \frac{g_s^2\,Q_1}{v\,Q_5}\,.
  \end{aligned}
 \label{eq:nsns}
\end{equation}
The range of validity of this description is again constrained by two conditions
\begin{equation}
  \begin{aligned}[m]
    \frac{v}{Q_1\,Q_5} & <  \left(\frac{g_s}{Q_5}\right)^2 \quad \text{(weak coupling)} \\
    \beta^2\,Q_5 & >  1 \quad (R_{\text{eff}}^2 > \alpha^\prime)~.
  \end{aligned}
 \label{eq:wkcouple}
\end{equation}

\paragraph{S-duality followed by T-duality: }
\label{sec:sthentads}

If the radius of the compact direction becomes much smaller than the string scale in \eqref{eq:nsns}, we need to go to a 
T-dual description of the above S-dual configuration.  
As already noticed in \cite{CH94}, the self-dual orbifold of $\ads{3}$ has the significant property that
if we apply the T-duality transformations \cite{tdualref} along the fiber generated by $\partial_\phi$, the type IIA
configuration thus obtained has  the same geometry. Technically,  this 
property can be easily seen by  the fact that the metric cross term $g_{t\phi}$ and the NS-NS 2-form potential
component, $B_{t\phi}= -\beta\,Q_5\sinh 2z$ are identical. 
So under their exchange
(T-duality), the geometry will not be modified. There is effectively just a rescaling of the original $\beta$
parameter, due to the transformation of the radius of the compact direction under T-duality. In other words, the
T-dual geometry is given in terms of \eqref{eq:nsns} but with a new parameter $\beta^\prime=(\beta\,Q_5)^{-1}$. This description 
is good when both conditions \eqref{eq:wkcouple} are violated. 
Notice that the self-dual radius is given by $\beta^2 = (Q_5)^{-1}$. 

\paragraph{T-duality: }
We now consider the T-dual configuration of \eqref{eq:rr}. This is natural when the effective radius of the compact
direction $\phi$ becomes smaller than the string scale. 
The T-dual configuration is 
\begin{equation}
  \begin{aligned}[m]
    g & = l^2\left(g_{\ads{2}} + g_{\sph{3}}\right) + \alpha^\prime\,{1\over g_s}\sqrt{\frac{v}{Q_1\,Q_5}}\,d\phi^2 +
    \alpha^\prime\sqrt{\frac{Q_1}{v\,Q_5}} dx^i\,dx_i ~,\\
    d\,C_3 &= Q_5 \left(\dvol\,\sph{3}\wedge d\phi\right)~, \\
    C_1 &=  \beta\,Q_5\,\sinh 2z\, dt ~, \\
    B_2 &= -\frac{1}{2\beta}\sinh 2z \,dt\wedge d\phi~,\\
    e^{-2\Phi} &= \frac{(\beta\,Q_5)^2}{g_s}\sqrt{\frac{v}{Q_1\,Q_5}}\,.
  \end{aligned}
 \label{eq:IIa}
\end{equation}
The geometry is given by $\ads{2}\times\sph{3}\times\sph{1}\times\bT^4$. The conformal boundary of this metric is the  
conformal boundary of $\ads{2}$, that is, two real lines. 

\paragraph{T-duality followed by M-theory lift: }
The above T-dual description is reliable at weak coupling,
\[
  \frac{(\beta\,Q_5)^2}{g_s}\sqrt{\frac{v}{Q_1\,Q_5}} > 1~.
\]
However at strong string coupling, an eleventh dimension $y$ opens up, giving
rise to the eleven dimensional configuration:
\begin{equation}
  \begin{aligned}[m]
    \frac{g}{l_p^2} &= \left((g_s\,\beta\,Q_5)^2\,\frac{Q_1\,Q_5}{v}\right)^{1/3}\left(-\cosh^2 2z\,dt^2 + dz^2 + g_{\sph{3}}\right) +
    \left(\frac{\beta\,v}{g_s^2\,Q_1}\right)^{2/3}\,d\phi^2 \\
    & + \left(\frac{\beta^2\,Q_1}{v\,g_s}\right)^{1/3}\,dx^idx_i + \left(g_s^2\,\frac{Q_1\,Q_5}{v}\,(\beta\,Q_5)^{-4}\right)\,
    \left(dy + \beta\,Q_5\,\sinh 2z\,dt\right)^2 \\
    F_4 &= dC_3 - dy\wedge H_3 = \left(Q_5\,\dvol\,\sph{3} - \beta^{-1}\dvol\,\ads{2}\wedge dy\right)\wedge d\phi
  \end{aligned}
 \label{eq:mtheory}
\end{equation}
This, again has the topology of an $\sph{1}$ fibration over $\ads{2}$ times $\sph{2}\times\sph{1}\times\bT^4$.

\subsection{Asymptotically flat construction}
\label{sec:flat}

It is interesting to study the asymptotically flat counterparts of the various asymptotically AdS solutions given in Sec \ref{sec:duals}. 
Our main motivation is that in the aymptotically flat constructions, these spacetimes acquire a brane interpretation, which can provide
some intuition concerning the dynamics of the system. We start with the D1-D5 system with momentum along the common direction, which leads to an 
$\ads{3} \times \sph{3}$ factor in the near horizon limit. In the asymptotically flat spacetime, the momentum is described by adding a third conserved 
charge, which leads us to add a third harmonic function which would source this charge. Our discussion is analogous to that for the non-dilatonic 
branes in \cite{cvetic}. We follow the same strategy but for the non-dilatonic D1-D5 system. Therefore, we consider the metric ansatz
\begin{multline}
  \frac{g}{\alpha^\prime} = (f_1\cdot f_5)^{-1/2}\left\{(W-2)dt^2 + W\,dz^2 - 2(W-1)\,dt\,dz\right\} + f_1^{1/2}\,f_5^{-1/2}\,dx^idx_i \\
  + (f_1\cdot f_5)^{1/2}\left(dr^2 + r^2\,g_{S^3}\right)\,,
 \label{eq:d1d5wave}
\end{multline}
where 
\[
  f_1 = 1 + \frac{\alpha^\prime\,g_s\,Q_1}{v\,r^2} \quad , \quad f_5 = 1 + \frac{\alpha^\prime\,g_s\,Q_5}{r^2} 
  \quad , \quad W = 1 + \frac{\alpha^\prime\,N}{v\,r^2}~,
\]
and $N$ could still depend on the string coupling constant $g_s$. Here, $z$ is an angular variable. 
The solution of type IIB equations of motion requires non-trivial dilaton and RR three-form field strength. These are given by
\begin{equation}
  \begin{aligned}[m]
    e^{2\Phi} &= g_s^2\,\frac{f_1}{f_5} ~, \\
    F_3 &= -g_s^{-1}\,df_1^{-1}\wedge dt\wedge dz + 2Q_5\dvol\,S^3~.
  \end{aligned}
 \label{eq:d1d5waveflux}
\end{equation}

The near horizon limit of such pp-waves propagating in non-dilatonic branes does not lead to the AdS space. Instead, we obtain the 
Kaigorodov spacetime (see \cite{cvetic} for details). Interestingly, in three dimensions, Kaigorodov spacetime is known to be equivalent 
to the extremal BTZ black hole. Therefore, taking the standard decoupling limit, keeping the momentum density of the wave fixed, we get the 
construction in \eqref{eq:5bh}, which was discussed in \cite{S98}. Then, after a very-near-horizon limit  \eqref{verynearhorizon} leads to 
a local description of the orbifold of $\ads{3}$ under discussion in this paper.

As is Sec \ref{sec:duals} it is interesting to perform duality transformations on this solution. 

\paragraph{S-duality: }
We now consider the S-dual of the configuration in  \eqref{eq:d1d5wave} and \eqref{eq:d1d5waveflux}. Given the brane interpretation of this 
type IIB configuration, it is clear that the S-dual configuration will be a F1-NS5 system with some momentum (wave)
propagating along the common direction. Indeed, the metric, dilaton and fluxes in the S-dual are given by:
\begin{equation}
  \begin{aligned}[m]
    e^{2\Phi} &= g_s^{-2}\,\frac{f_5}{f_1} ~, \\
    H_3 &= g_s^{-1}\,df_1^{-1}\wedge dt\wedge dz - 2Q_5\dvol\,S^3~, \\
    \frac{g}{\alpha^\prime} & = g_s^{-1}\left(f_1^{-1}\left\{(W-2)dt^2 + W\,dz^2 - 2(W-1)\,dt\,dz\right\} + dx^idx_i
    + f_5\,\left(dr^2 + r^2\,g_{S^3}\right)\right)~.
  \end{aligned}
 \label{eq:f1ns5wave}
\end{equation}

\paragraph{S-duality followed by T-duality: }
The T-dual of this S-dual configuration \eqref{eq:f1ns5wave} along the circle parameterized by $z$ explains  the self-duality property discussed in the asymptotically AdS discussion in Sec. 
\ref{sec:sthentads}. Indeed, such a T-duality transformation
interchanges the wave with the fundamental string, so that the resulting configuration has no modification
in the geometry and fluxes. The configuration is given exactly by the set of equations \eqref{eq:f1ns5wave} interchanging the role
played by the harmonic functions $f_1$ and $W$, up to rescaling of coordinates.

\paragraph{T-duality: }
The T-dual description of the configuration \eqref{eq:d1d5wave}-\eqref{eq:d1d5waveflux} along the circle parameterized
by $z$  gives rise to a D0-D4 system with fundamental strings winding around the dual circle, in which all the D-branes
are delocalized in the T-dual circle. The dilaton and fluxes describing this configuration are given by
\begin{equation}
  \begin{aligned}[m]
    e^{2\Phi} &= g_s^2\,W^{-1}\,f_1^{3/2}\,f_5^{-1/2}~, \\
    F_2 &= -g_s^{-1}\,df_1^{-1}\wedge dt ~, \\
    F_4 &= 2Q_5\,\dvol\,S^3\wedge dz~, \\
    H_3 &= dW^{-1}\wedge dt\wedge dz~,
  \end{aligned}
 \label{eq:d0d4f1flux}
\end{equation}
whereas the metric is
\begin{multline}
  \frac{g}{\alpha^\prime} = f_1^{1/2}\,f_5^{-1/2}\,dx^idx_i + (f_1\cdot f_5)^{1/2}\,\left(dr^2 + r^2\,g_{S^3}\right) \\
  - (f_1\cdot f_5)^{-1/2}\,W^{-1}\,dt^2 + (f_1\cdot f_5)^{1/2}\,W^{-1}\,dz^2~.
 \label{eq:d0d4f1}
\end{multline}

\paragraph{T-duality followed by M-theory lift: }
Finally, we could consider the strong coupling description of the T-dual configuration, where the eleventh dimension $y$ opens
up. This is expected to describe an M2-M5 system sharing one direction with momentum propagating along it. The eleven dimensional
configuration is described by the metric
\begin{multline}
  \frac{g}{l_p^2} = (g_s\,W)^{-2/3}\,f_5^{2/3}\,dy^2 + (g_s\,W)^{1/3}\,f_5^{-1/3}\,g_s^{-1}\,dx^idx_i + (g_s\,W)^{1/3}\,f_5^{2/3}\,g_s^{-1}
  \left(dr^2 + r^2\,g_{S^3}\right) \\
  + (g_s\,W)^{-2/3}\,f_5^{-1/3}\left\{(f_1-2)\,dt^2 + 2(f_1-1)\,dtdz + f_1\,g_s^2\,dz^2\right\}~,
 \label{eq:m2m5wave}
\end{multline}
and the four-form field strength
\begin{equation}
  F_4 = dW^{-1}\wedge dt\wedge dz\wedge dy - 2Q_5\dvol\,S^3\wedge dy~.
 \label{eq:m2m5waveflux}
\end{equation}

\subsection{Towards a matrix model description}

The string dualities described in Sec.~\ref{sec:duals} and the related asymptotically flat brane constructions in Sec.~\ref{sec:flat} relate our 
spacetime to matrix model descriptions. First, note again, as in Sec.~\ref{penroselike} that our orbifold has a description in terms of a Penrose-like 
limit of the D1-D5 string wound on a circle with left-moving momentum, and, correspondingly, in terms of a DLCQ of the D1-D5 CFT. 
Let the charges of these branes and momenta be $(Q_1, Q_5, n)$.

T-duality along the common circle direction produces a D0-D4-F1 system in IIA theory with charges $(Q_1, Q_5, n)$, where now $n$ 
measures the winding of the F1s along the T-dual circle.  When the charge $Q_1$ is very large the entire configuration should be 
describable in a $U(Q_1)$ matrix quantum mechanics associated with the D0s. Equivalently, the lift to M-theory of this configuration 
describes a M2-M5-p system with n M2s intersecting the $Q5$ M5s on the 11th circle with $Q1$ units of momentum. In the limit 
of large Q1, this background can be described as a state in the BFSS matrix model \cite{BFSS} which is again a matrix mechanics.

The S-dual of our orbifold in Secs.~\ref{sec:duals} and \ref{sec:flat} arises as Penrose-like limit of of the F1-NS5 system in IIB string theory 
with momentum along the common direction and charges $(Q1,Q5,n)$.  The Penrose-like limit corresponds to a low-energy limit 
in a sector where the momentum is going to infinity.  This is precisely a scenario in which the $(0,2)$ worldvolume theory on the NS5-branes 
reduces to a quantum mechanics since the 5-brane is wrapped on a 4-torus and a circle.  What is more, the large momentum $n$ implies a 
DLCQ of the $(0,2)$ theory \cite{dlcq} and is thus related to quantum mechanics on the $n$ instanton moduli space of a $U(Q_5)$ gauge theory 
in four dimensions. The fundamental strings will be represented as a certain state of this quantum mechanics. This again leads to a description 
in a matrix quantum mechanics.  Interestingly, as we showed earlier, the F1-NS5-p system is self-dual under T-duality. This self-duality must 
manifest itself in the matrix model description. We can also exchange the M-theory circle with the circle in the IIA NS5-F1-p solution.  
This gives a M5-M2-p system in M-theory with n units of momentum along the common direction of the M5 and the M2.  
As described in Sec.~\ref{penroselike} the Penrose-like limit leading to our orbifold geometry arises is a large $n$ limit.  
Thus our background will arise as a state in the BFSS matrix model describing longitudinal M5 and M2 branes.

\section{Conclusion}

In this paper we have examined an interesting orbifold of $\ads{3}$  by boosts and established many intriguing properties and connections with other corners of string theory.  The boundary of our space contains two ``null cylinders'' and we established that a DLCQ of the D1-D5 string appears in the holographic description.   We also showed that from several perspectives a matrix model is involved in holography, a fact which is particularly interesting because our orbifold is the universal geometry in the vicinity of the finite area extremal black holes of string theory in 4d and 5d and also of the BTZ black holes.    
As we discussed, it is natural to compactify our orbifold to two dimensions, and so we can also regard these comments as indicating a relation between certain 2d string theories and corresponding matrix models. 
This seems intriguing, although there is no direct connection between this observation and  the $c=1$ matrix model description of 2d bosonic string theory.  Note also, that the various S and T dualities of our orbifold imply corresponding dualities between the various matrix model and DLCQ holographic descriptions of the spacetime.

A productive methodology in the study of pp-waves was to implemented the Penrose limit as an operation in the dual to the full AdS spacetime thereby isolating a sector of the full CFT  that described just  the pp-wave.  A useful way of making progress here might be to consider an analogous operation focusing on a sector of the dual to the BTZ hole hole which has also been extensively studied.  It would also be interesting to consider our orbifold as a solution in the Chern-Simons description of 3d gravity and to explore excitations of this background such conical defects which arise from additional orbifolding.   String theory on the orbifold can also be studied by explicitly constructing orbifolds of the WZW model description $\ads{3} \times S^3$.

Our spacetime does not have a natural Euclidean continuation, hence its interest as a laboratory for the study of time dependent string theory.   From many perspectives we saw that important properties of the spacetime and its dual are controlled by the structure of the complexified manifold constructed by continuing all the coordinates to complex values.   For example, excursions in the complex coordinate plane transported real points in the spacetime to other real locations, and a singularity at an imaginary radial coordinate controlled the structure of normalizable wave solutions.   This is reminiscent of the importance of the analytic structure and geodesics in the complex coordinate plane for 
 recent investigations of ``seeing holographically behind a horizon'' \cite{BTZholog}.    A general lesson to be learned from all of these works might well be that even in the absence of a Euclidean continuation for a general time dependent universe, physical properties are nevertheless controlled by the analytic structure of complexified spacetime.


\section*{Acknowledgements}
We are grateful to Micha Berkooz,  Jan de Boer, Ben Craps,  Jose Figueroa-O' Farrill, Eric Gimon, Esko Keski-Vakkuri, Matt Kleban, Per Kraus, Tommy Levi, Liat Maoz, Emil Martinec, Djordje Minic, 
Simon Ross, Kostas Skenderis, Andy Strominger and Jan Troost for interesting conversations at various stages in this project.   
Part of this work was completed at the Aspen Center for Physics.   
JS would like to thank the University of Chicago, the Institute for Advanced Studies in Princeton and the 
Perimeter Institute for kind hospitality during the different stages involved in this project. 
JS was supported by a Marie Curie Fellowship of the European
Community programme ``Improving the Human Research Potential and the Socio-Economic Knowledge Base'' under the
contract number HPMF-CT-2000-00480, during the initial stages of this project, by the Phil Zacharia fellowship
from January to August in 2003.  The travelling budget  of JS was also supported in part by a grant from the United States--Israel
Binational Science Foundation (BSF), the European RTN network HPRN-CT-2000-00122 and by Minerva, during the
initial stages of this project. JS would also like to thank the IRF Centers of Excellence program. AN was supported in part by Stiching FOM.  
Work on this project at Penn was supported by the DOE under grant DE-FG02-95ER40893 and by the NSF under grant PHY-0331728.

\appendix
\section{Two point functions in momentum space}
\label{sec:fourier}
  The Fourier transform of $G_{++}$ is:
\begin{equation}
\tilde{G}_{++}(\omega, p) = 
\int_{-\infty}^\infty dt  \int_{-\infty}^\infty d\phi \, \,  e^{i \omega t} e^{i p \phi} G_{++}(t,  \phi)
=  {i \over 2\pi R} \, h(\omega) \, g(p)
\end{equation}
where $h$ and $g$ are the Fourier transforms of the $t$ and $\phi$ dependent parts respectively.
(Since $G_{++}$ is periodic in $\phi$ the integral over the latter will yield a delta function that forces $p$ to be an integer.)   We will do this transform by analytic continuation of the integral
\begin{equation}
  \int_0^\infty e^{-\alpha x} \,  (\sinh\beta z)^\gamma \, dx = {1 \over 2^{\gamma +1} \beta} \, 
  B(\alpha/2\beta - \gamma/2, \gamma +1)
\label{masterintegral}
\end{equation}
where $B$ is the beta function
\begin{equation}
B(x,y) = {\Gamma(x) \Gamma(y) \over \Gamma(x+y)}
\end{equation}
and the integral assumes ${\rm Re}(\beta) >0$, ${\rm Re}(\gamma) > -1$, ${\rm Re}(\alpha) > {\rm Re}(\beta \gamma)$.

Consider the integral over $\phi$ first.  Let
\begin{equation}
g_n(p) = \int_{-\infty}^\infty d\phi \, \,  e^{i p \phi} [\sinh(\phi + 2\pi n)]^{-2h_+}
\end{equation}
Since the range of $\phi$ is unbounded in the integral we can shift the integrand getting
\begin{equation}
g_n(p) = e^{-i2\pi p n} \int_{-\infty}^\infty d\phi \, \, e^{i p \phi} [\sinh(\phi)]^{-2h_+}
= e^{-i2\pi p n} g_0(p)
\end{equation}
Thus 
\begin{equation}
g(p) = \sum_{n=-\infty}^{n = \infty} e^{-i2\pi p n} g_0(p)
= g_0(p) \, \delta(p -m)  ~~~~;~~~~   m \in Z
\end{equation}

To compute $g_0(p)$ we split the integral over positive and negative $\phi$ using $\sinh(-\phi) = -\sinh(\phi)$.  This gives
\begin{eqnarray}
g_0(p) &=&
\int_{-\infty}^\infty d\phi \, \, e^{ip\phi} (\sinh\phi)^{-2h_+} \\
&=& \int_0^\infty d\phi \, \,  e^{ip\phi} (\sinh\phi)^{-2h_+} 
+ (-1)^{2h_+} \,  \int_0^\infty d\phi \, \,  e^{-ip\phi} (\sinh\phi)^{-2h_+}  
\end{eqnarray}
We do these integrals by using (\ref{masterintegral}) and analytically continuing the parameters in the equation above as
\begin{equation}
\gamma = -2 h_+ + a ~~~;~~~
\beta = 1 ~~~;~~~
\alpha = -ip + b
\end{equation}
with $a$ and $b$ large enough so that the conditions for the validity of (\ref{masterintegral}) are satisfied.   We then apply (\ref{masterintegral}) and continue $a,b \to 0$. This gives
\begin{equation}
g_0(m) =
{\Gamma(1 - 2h_+) \over 2^{1-2h_+} }\left[
{\Gamma(-i m/2 + h_+) \over \Gamma(-i m/2 + h_-) }
+ (-1)^{-2h_+} {\Gamma(i m/2 + h_+) \over \Gamma(i m/2 + h_-) }
\right]
\end{equation}
Similarly, the Fourier transform of the time part of $G_{++}$ gives
\begin{equation}
h(\omega) =
{\Gamma(1 - 2h_+) \over  (2i)^{1-2h_+} }\left[
{\Gamma(- \omega /2 + h_+) \over \Gamma(-\omega/2 + h_-) }
+ (-1)^{-2h_+} {\Gamma(\omega/2 + h_+) \over \Gamma(\omega/2 + h_-) }
\right]
\end{equation}
Multiplying everything together we find
\begin{multline}
  \tilde{G}_{++}(\omega,m) =  {i \over 2\pi R} \, (-1)^{-2h_+}\frac{1}{4}\,\left(\frac{i}{2}\right)^{2h_+-1} \frac{1}{2^{2h_+}}\,\left(\Gamma (1 - 2h_+)\right)^2\,
  \frac{\Gamma (h_+ - im/2)}{\Gamma (h_- - im/2)}\,\frac{\Gamma (-\omega/2 + h_+)}{\Gamma (-\omega/2 + h_-)} \\
\times \, 
  \left(e^{i\theta} + (-1)^{-2h_+}\right)\left\{ \frac{\sin (\pi(\omega/2 + h_-))}{\sin (\pi(\omega/2+h_+))}+ (-1)^{-2h_+}\right\}~.
 \label{fourierfinal2}
\end{multline}
where we  used the relation
\begin{equation}
  \Gamma(z) \Gamma(1-z) =\pi \csc \pi z
\label{asadlove}
\end{equation}
several times and the definition
\begin{equation}
 {\csc\Bigl( \pi ( h_++{im \over 2}) \Bigr) \over \csc\Bigl( \pi ( h_- +{im \over 2}) \Bigr)}=e^{i \theta} \, .
\end{equation}


\end{document}